\documentclass[aps,prd,twocolumn,superscriptaddress,showpacs,nofootinbib]{revtex4}

\usepackage{graphicx}

\newcommand{\povo}{Dipartimento di Fisica, Universit\`{a} di Trento,
and I.N.F.N., Gruppo di Trento,
38050 Povo (TN), Italy}

\newcommand{\figr}[1]{Fig.\ref{#1}}

\newcommand{\eqr}[1]{Eqn.\ref{#1}}
\newcommand{\sref}[1]{Sec. \ref{#1}}
\newcommand{\aref}[1]{App. \ref{#1}}

\newcommand{\tabr}[1]{Table \ref{#1}}

\newcommand{\beq}[1]{\begin{equation} \label{#1}}
\newcommand{\eeq}{\end{equation}}
\begin{document}

\preprint{preprint}

\title{Thermal gradient-induced forces on geodetic reference masses for LISA}

\author{L. Carbone}
\affiliation{\povo}
\author{A. Cavalleri}
\affiliation{Centro Fisica degli Stati Aggregati, 38050 Povo (TN), Italy}
\author{G. Ciani}
\affiliation{\povo}
\author{R. Dolesi}
\affiliation{\povo}
\author{M. Hueller}
\affiliation{\povo}
\author{D. Tombolato}
\affiliation{\povo}
\author{S. Vitale}
\affiliation{\povo}
\author{W. J. Weber}
\affiliation{\povo}

\date{\today}

\begin{abstract}
The low frequency sensitivity of space-borne gravitational wave
observatories will depend critically on the geodetic purity of the
trajectories of orbiting test masses.  Fluctuations in the
temperature difference across the enclosure surrounding the
free-falling test mass can produce noisy forces through several
processes, including the radiometric effect, radiation pressure, and
outgassing.  We present here a detailed experimental investigation
of thermal gradient-induced forces for the LISA gravitational wave
mission and the LISA Pathfinder, employing high resolution torsion
pendulum measurements of the torque on a LISA-like test mass
suspended inside a prototype of the LISA gravitational reference
sensor that will surround the test mass in orbit.  The measurement
campaign, accompanied by numerical simulations of the radiometric
and radiation pressure effects, allows a more accurate and
representative characterization of thermal-gradient forces in the
specific geometry and environment relevant to LISA free-fall.  The
pressure dependence of the measured torques allows clear
identification of the radiometric effect, in quantitative agreement
with the model developed.  In the limit of zero gas pressure, the
measurements are most likely dominated by outgassing, but at a low
level that does not threaten the LISA sensitivity goals.

\end{abstract}


\maketitle

\section{Introduction}

LISA, the Laser Interferometer Space Antenna \cite{bender:LISA}, is
a ESA-NASA mission to create the first spaceborne interferometric
observatory of gravitational waves, in the frequency range between
30~$\mu$Hz and 1~Hz. LISA should permit not only {\it detection},
but detailed, high resolution, long measurement time {\it
observation} of gravitational wave signals, opening the possibility
for important discovery in fundamental physics, astrophysics, and
cosmology.

LISA consists of three identical spacecraft separated by $5 \times
10^6$ km, forming a nearly equilateral triangle that orbits the sun
at a distance of 1~AU. Inside each spacecraft, there are two
``free-falling'' cubic test-masses (TM) in nominally geodesic
motion. Gravitational waves are detected by an interferometric
measurement of the relative change in the distances between distant
free-falling TM caused by the gravitational wave metric
perturbation. LISA aims at a gravitational wave strain noise floor
of $4 \times 10^{-21} \mathrm{/Hz^{1/2}}$ near 3~mHz, increasing as
$1/f^2$ at lower frequencies. Performance at these lower frequencies
is limited by the purity of the TM geodetic motion and requires
perfect free-fall, along the sensitive interferometry axis (referred
as the $x$ axis here), to within an acceleration noise of $3 \times
10^{-15} \mathrm{m/s^2/Hz^{1/2}}$ ($3\: \mathrm{fm/s^2/Hz^{1/2}}$).
For the 2~kg TM foreseen for LISA, this is equivalent to a force
noise of 6~fN/Hz$^{1/2}$. While placing test particles in nearly
perfect geodetic orbits inside of a controlled, co-orbiting
spacecraft is essential to various space gravitational experiments,
the extremely low force noise requirement and low measurement
frequencies make LISA even more demanding in this respect than other
previous (GPB \cite{GPB}) and future (STEP \cite{STEP}, Microscope
\cite{Microscope}) missions.

While the spacecraft shields the TM from environmental disturbances
such as solar radiation pressure, it is itself a leading source of
disturbing force acting on the TM. Achieving the required extremely
low level of spurious acceleration requires identification and
suppression of all interactions that can compete with gravity in
defining the trajectory of the particle. The disturbances identified
thus far have been collected in an overall noise model
\cite{dolesi:sensor, stebbins:errors, bonnie:noise}that has been
used to optimize the LISA design.

\begin{figure}[t]
\includegraphics[width=85mm]{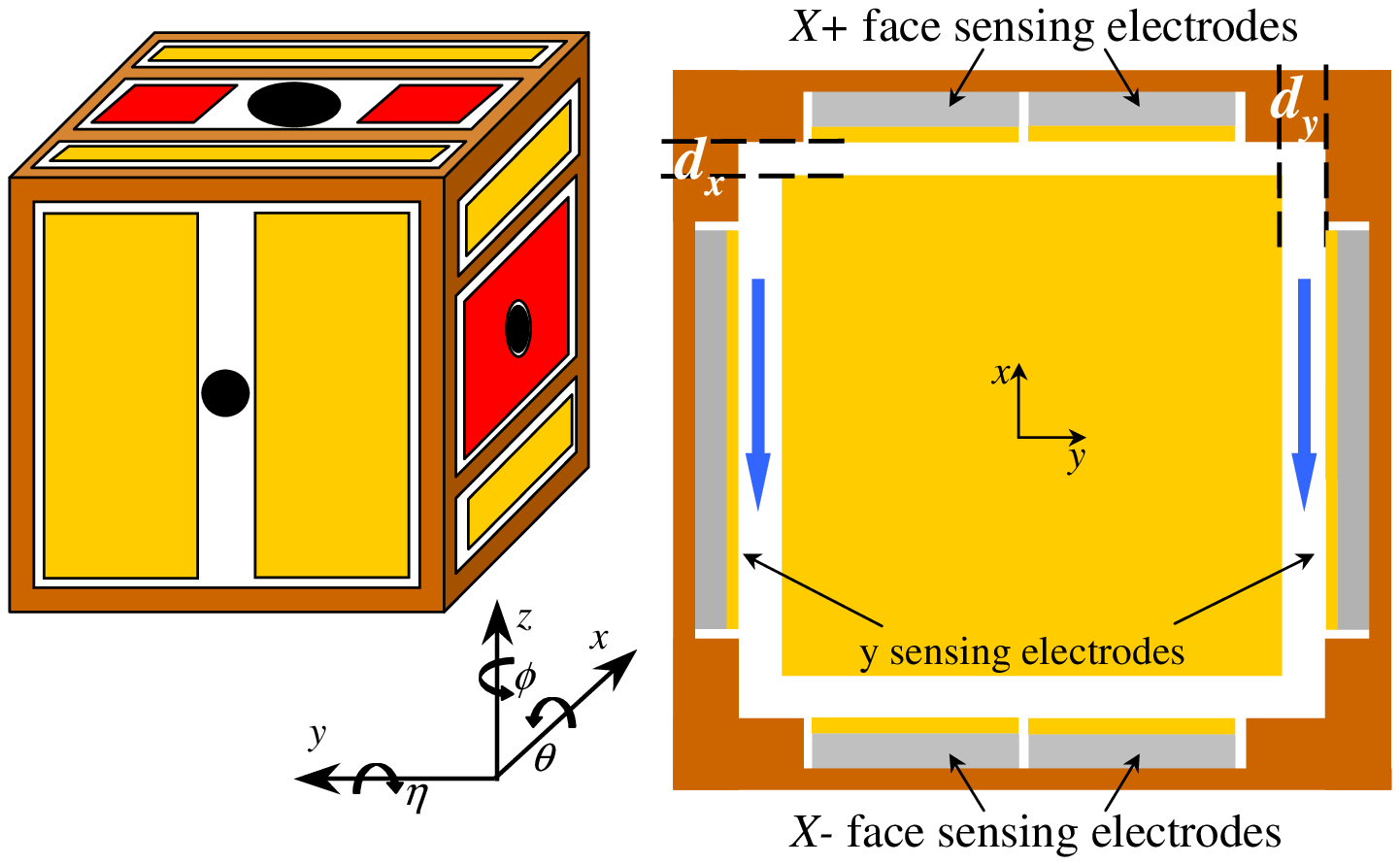}
\caption{Schematic of the six degree of freedom capacitive position
sensor for LISA Pathfinder and LISA. The sensor electrodes are
obtained by Au-coating of an insulating substrate, mounted on an
external metallic frame. At left, sensing electrodes are in orange,
sensor bias injection electrode are red, and the supporting housing
is brown. The black circles represent holes to allow access for
laser light and for a TM caging mechanism during launch. The $x$
axis represents the sensitive LISA interferometry axis relevant for
force noise. A internal section of the GRS is shown, with the blue
arrows representing the path linking the two $X$ faces of the TM for
gas and thermal photons. The TM-electrode gaps on the three axes are
$d_x =4\:$ mm, $d_y =2.9\:$ mm and $d_z =3.5\:$ mm.} \label{sens}
\end{figure}

Particular attention has been dedicated to the interaction between
the TM and the position sensor that surrounds the TM and is used to
guide precision thrusters that keep the spacecraft centered around
the free-falling TMs. As such, the design of the TM and position
sensor -- referred to together as the ``gravitational reference
sensor'' or GRS -- is critical to achieving the LISA force noise
goals. The current GRS is a 46 mm cubic TM of Au-Pt (2~kg)
surrounded by an array of conducting electrode surfaces used for a
capacitive position readout and force actuation scheme (see
\figr{sens} and \cite{weber:sensor,dolesi:sensor}). The TM material
is chosen for low magnetic susceptibility and residual moment, to
minimize interaction with magnetic field noise \cite{mauro:mag}. To
reduce short range stray electrostatic effects, the distance
(``gaps'') between the TM and the electrodes are relatively large,
4~mm on the interferometry axis, and all conducting surfaces are
Au-coated to provide electrostatic homogeneity. To limit temperature
differences across the TM, and thus also the temperature-gradient
related forces which are the subject of this article, the sensor
housing is made of a high thermal conductivity composite structure
of molybdenum and sapphire.

The unprecedentedly small level of force noise needed for LISA
requires measurement of the small disturbing forces in order to
verify the feasibility of the gravitational wave sensitivity goals.
Given the possibility of force noise sources that are not accurately
modelled or perhaps not even previously identified, these tests
should, as much as possible, use the final flight hardware
configuration, reflecting not only the nominal design materials and
geometry, but also the relevant machining, surface finishing,
impurities, and cleanliness. Such testing is being pursued both in
space, where the LISA Pathfinder mission will perform a dedicated
flight test of the full LISA drag-free control system
\cite{vitale:nuc:phys, vitale:LTP}, and on ground, where torsion
pendulum dynamometers can characterize the force noise generated
inside the GRS \cite{hueller:pend}. A recent torsion pendulum study
of the GRS capacitive sensor to be flown aboard LISA Pathfinder has
demonstrated the absence of unknown surface forces to within roughly
an order of magnitude of the LISA goal at 3~mHz
\cite{prd:force_noise}. Other studies have measured important
electrostatic and elastic coupling
effects\cite{carbone:prl,carbone:char,carbone:thesis,hueller:upperlimits,seattle}.
Ongoing work aims at extending these studies to lower force noise
levels, lower frequencies, and to even more representative hardware
configurations.

This paper describes a detailed study of the specific class of force
noise induced by thermal gradients.  Thermal gradient-induced forces
have been modeled as part of the acceleration noise budget for LISA
\cite{bonnie:noise, stebbins:errors,
dolesi:sensor,nobili,nobili_astr} and have been the subject of
preliminary experimental investigations
\cite{carbone:char,carbone:thesis,pollack:thermal}. A temperature
difference between the sensor surfaces on opposing sides of the TM
converts into a net force through at least three mechanisms: the
radiometric effect, differential radiation pressure, and temperature
dependent outgassing. Broadly speaking, the radiometric and
radiation pressure forces arise in the temperature dependence of the
momentum transferred to a TM face by impacts of, respectively,
residual gas molecules and thermal photons. If, at the LISA
operating temperature (293~K) and pressure (10$^{-5}$~Pa), we apply
a simplified ``infinite plate'' model - to be discussed, along with
its limitations and a more accurate simulation, in
\sref{estim:theor} -- the radiometric and radiation pressure effects
can be estimated to produce $\frac{dF}{d\Delta T}$, a net force per
degree of temperature difference, of roughly 20 and 30~pN/K,
respectively, always pointing from the hot side of the sensor
towards the cold. A third effect, more difficult to model and
dependent on surface cleanliness, arises in the temperature
dependence of outgassing from the sensor surfaces and the net
momentum thus imparted on the TM by desorbed molecules in the
presence of a thermal gradient. As will be discussed in the
following section, this effect has been expected to contribute
roughly as much as the other two effects. The three effects respond
coherently to the same temperature gradient and sum to give a net
temperature-difference response of roughly $\frac{dF}{d\Delta T}
\approx$ 100 pN/K.

This estimated ``transfer function'' sets the tolerable level of
temperature difference fluctuations. In order to hold the thermal
gradient-induced force noise to roughly $10\%$ of the overall force
noise budget, the fluctuations in temperature difference across the
GRS must be less than 10$^{-5}$ K/Hz$^{1/2}$. This level of
temperature difference stability, which motivated the use of high
thermal conductivity materials in the sensor, looks feasible but is
challenging at the lowest frequencies, where the passive thermal
filtering of solar radiation intensity fluctuations becomes
progressively less effective\cite{Bender,Merko}.

The article aims at a characterization, through experiment and
simulation, of the thermal gradient transfer function
$\frac{dF}{d\Delta T}$ relevant to LISA free-fall. This study is
important to free-fall for LISA and other precision force
measurements for several reasons. First, while the rough estimates
discussed above for the radiation pressure and radiometric effects
are readily obtained from simple formulas, they are based on an
infinite plate model, which is not particularly accurate for the
geometry of the proposed GRS, where the dimensions of the TM are
only roughly 10 times larger than the TM-sensor gaps. As will be
seen from the results of the simulations presented in
\aref{radiom:app} and \aref{rapress:app}, these corrections, based
on the geometry and, in the case of radiation pressure, surface
reflectivity properties of the sensor, have a quantitative impact on
the estimates of these noise sources for LISA and on the
measurements presented here.

An even more important motivation for an experimental investigation
lies in the large uncertainties with which we can estimate the
outgassing effect, which will depend not only on the sensor
geometry, but also on the amount, geometrical distribution, chemical
nature, and history of the surface contamination inside the sensor.
The amount to which the purity of free-fall is threatened by
outgassing can only be reliably estimated through measurement with
representative sensor hardware. The same argument can be applied to
any other possible unmodelled or unidentified thermal related
disturbance.

It should be noted that knowing the impact of thermal gradients can
have an impact on the overall LISA experiment design, particularly
on the amount of thermal filtering needed, or even whether active
temperature control is necessary. Additionally, the measurement of
$\frac{dF}{d\Delta T}$ can be repeated in-flight during the LISA and
LISA Pathfinder missions, a ``calibration'' that, combined with the
appropriate thermometer readings, will allow a correlation analysis
of the noise and even a subtraction of the disturbance from the
gravitational wave time series.

The experimental investigation was performed by means of a torsion
pendulum, with a thin fiber suspending a hollow LISA-like test mass
on-axis with its center of mass, inside a GRS prototype. Two
prototype sensors were tested, the current LISA and LISA Pathfinder
design showed in \figr{sens}, with 4, 3.5, and 2.9~mm gaps, and an
earlier prototype with smaller gaps, 2~mm on all axes (these are
described in Sec. \ref{GRS:prototypes}). The geometry of the
pendulum is sensitive only to torques rather than the net forces
that are relevant to $\frac{dF}{d\Delta T}$ along the LISA
interferometer $x$ axis. As such, the basic measurement technique,
described in detail in \sref{experiment}, will be to apply a
``rotational temperature gradient'' such as to excite a measurable
torque that is proportional to the applied heat and, through
modelling of the different effects, to the transfer function
$\frac{dF}{d\Delta T}$. The measurements, of excited torque as a
function of the temperature gradient pattern and presented in
\sref{exp_results_disc}, are performed for various sensor average
temperatures and for a range of pressures, which allows a separation
of the radiometric and radiation pressure effects from their known
functional dependencies.

The pendulum configuration and the measurement of a torque (rather
than a force) signal necessarily complicates the experimental
analysis, demanding an interpolation of the relevant sensor
temperature distribution from a limited array of thermometers and
comparison with the torques calculated from the radiometric and
radiation pressure models. In addition to complicating the analysis,
the torque measurement imposed by the pendulum configuration means
that the measurement will be blind to potential outgassing effects
that act directly on the center of the TM faces, as there will be no
``effective armlength'' to convert such forces into measurable
torques. However, in spite of these analysis difficulties, discussed
in \sref{experiment}, and the limitations of the torque measurement,
to be discussed along with the study results in \sref{conclusions},
relevant conclusions can be drawn concerning the role of all
relevant contributions to the overall thermal gradient effect for
LISA free-fall.

\section{Modeling of thermal gradient-induced forces in the LISA GRS}
\label{estim:theor}

In this section we describe the three main physical processes that
convert fluctuating temperature gradients into noisy forces on the
LISA TM. For the radiometric and radiation pressure effects, we
present both the simplified models mentioned in the introduction and
the results from numerical simulations using realistic sensor
dimensions and surface properties.  A thorough explanation of the
simulations is given in Appendixes \ref{radiom:app} and
\ref{rapress:app}.

In the discussion that follows, we assume the average GRS
temperature to be $T_{0}$, with $T(\mathbf{r})$ the temperature
distribution on the internal surface of the position sensor, facing
the TM at a position $\mathbf{r}$. Given the high thermal
conductivity of the sensor materials and the small expected thermal
loads, we consider small temperature differences $\delta
T(\mathbf{r})\equiv\left(T(\mathbf{r})- T_{0}\right)\ll T_{0}$.
Additionally, as the thermal conductance across the TM is much
greater than that between TM and sensor, at least at a pressure of
10$^{-5}$ Pa, the calculations assume the TM to be isothermal at
$T_0$. Finally, to connect the results with the transfer function
$\frac{dF}{d\Delta T}$ relevant to LISA and discussed in the
introduction, we define the average temperature difference $\Delta
T_x \equiv \overline{T}_{x_{-}} -\overline{T}_{x_{+}}$
\footnote[1]{In this sign convention, $\Delta T_x$ positive will
create a positive $F_x$, and thus the various contributions to
$\frac{dF_x}{d \left( {\Delta T_x} \right)}$ will be positive.}
between the average temperature of the two opposing inner sensor
surfaces orthogonal to the $x$ axis, with the positive and negative
$X$ surfaces defined $X_{+}$ and $X_{-}$ as in \figr{sens}. While
for LISA only the force component along the sensitive $x$ axis,
$F_x$, is relevant, here we also calculate the $z$ component of the
torque, $N_z$, that coincides with the torsion pendulum fiber axis
and is thus relevant to the measurements presented in this article.

\subsection{Radiometer effect}\label{radiom:theor}
In gas systems where the mean free path is long compared to the
container dimensions -- such as for the LISA GRS at $10^{-5}$ Pa --
the hydrostatic equilibrium condition of pressure uniformity, which
is valid in the dense gas limit, is no longer relevant.  In the
presence of a temperature gradient, with the equilibrium condition
that there be no net flux of molecules between hot and cold zones, a
stable pressure difference is established.  The radiometer effect
refers to this steady state pressure difference that, for LISA,
arises in the net difference in momentum transferred to opposing
faces of the TM in the presence of a temperature gradient.

\begin{figure}
\includegraphics[width=100mm]{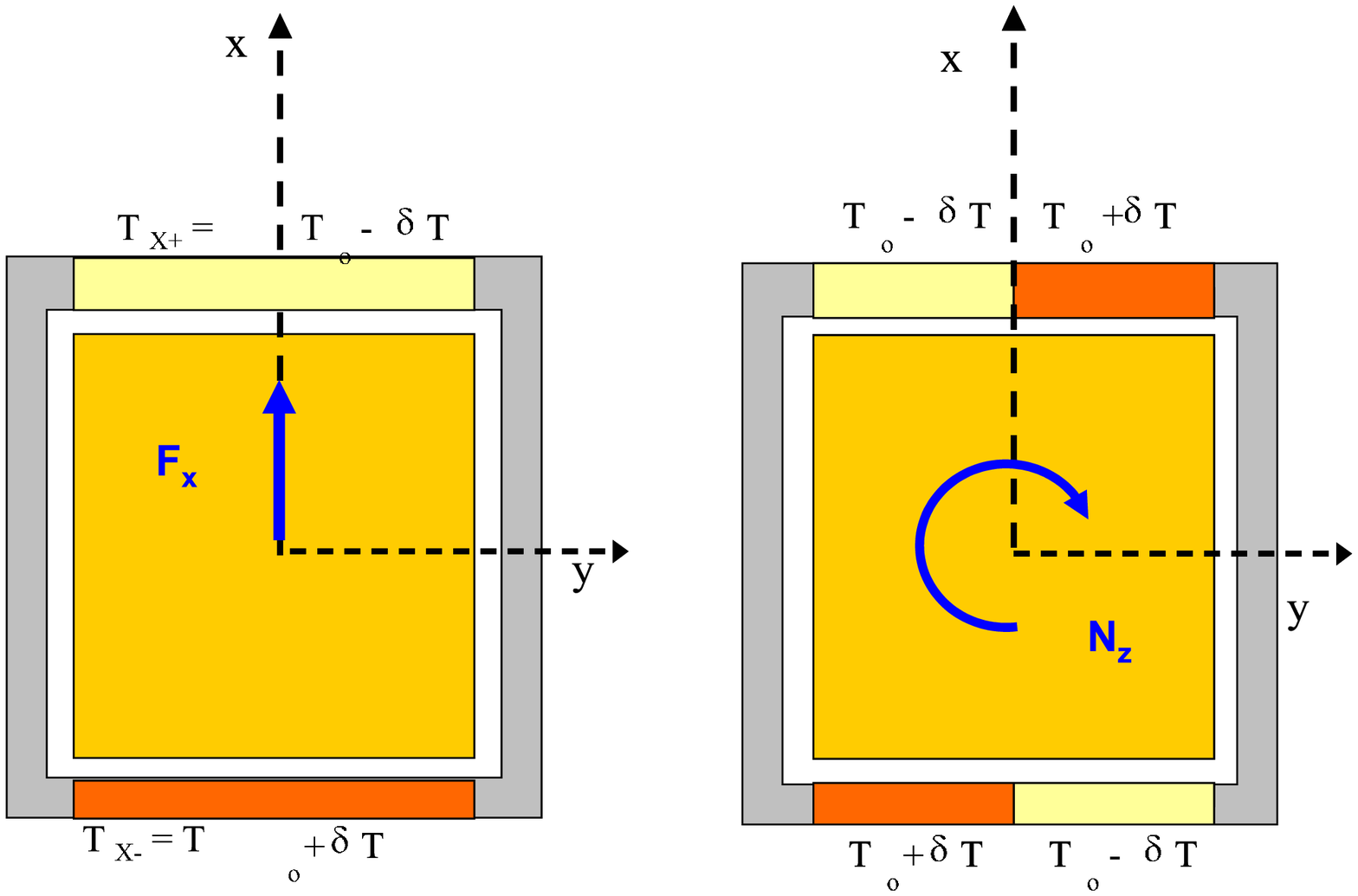}
\caption{Idealized temperature gradient configurations. The orange
square represents a section of a cubic TM of size L at uniform
temperature $T_0$. Yellow and red parts represent domains of uniform
temperature on the inner surface of the electrode housing, as
indicated next to them. Left panel: because of radiometer effects
(and, as shown in \sref{rapress:theor}, radiation pressure), the
illustrated temperature pattern is expected to induce a net force
$F_x$. Right panel: the illustrated temperature pattern is expected
to induce a torque $N_z$ around the TM $z$ axis. In the simplified
infinite plate model of the radiometer and radiation pressure
effects, the gray parts of the electrode housing do not contribute
to $F_x$, even in the presence of a temperature gradient (see
\eqr{Force:radiom}). The gray part can, in this simplified model,
contribute to $N_z$, in the event of a thermal gradient, through the
forces that they cause orthogonally to the TM faces in front of them
(as in \eqr{radiom:torqtheo}).} \label{tempgrad:ideal}
\end{figure}

Radiometric forces can be estimated with the transpiration theory
for a gas in the free-molecular regime \cite{Roth:book,Wu,Wu2},
which considers that the momentum distribution of molecules arriving
on a surface element depends on the temperature of the last surface
from which they departed, and will thus be affected both by the
system's specific temperature distribution and geometry. We consider
here the pressure $P_{pl}$ between two parallel plates, ``infinite''
in the sense that their separation is negligible compared with their
linear extent.  With the two plates at different uniform temperature
$T_{1}$ and $T_{2}$ and in equilibrium with a particle reservoir at
pressure $P$ and temperature $T_0$, the approximate solution is

\begin{equation}
P_{pl}=\frac{P}{2}\left(\sqrt{\frac{T_{1}}{T_{0}}}+
\sqrt{\frac{T_{2}}{T_{0}}}\right). \label{Fradiomparplate}
\end{equation}
This result holds even in the case that the molecules are not fully
``thermalized'' upon colliding with the surfaces and is only
slightly modified if the effective thermalization factor is
different between the two surfaces \cite{Wu}. In the limit of
TM-sensor gaps that are much smaller than the effective spatial
extent of the temperature perturbations, we apply this formula to
obtain the force $dF_{R}(\mathbf{r})$ (``R'' for radiometric) which
acts normally on TM surface element $d\mathbf{s}$ at position
$\mathbf{r}$ :
\begin{equation}
 dF_{R}(\mathbf{r})\simeq \frac{P}{2}
\left(1+\sqrt{\frac{T(\mathbf{r})}{T_{0}}}\right)d\mathbf{s}\simeq
Pd\mathbf{s}+\frac{P}{4}\frac{\delta T(\mathbf{r})}{T_{0}}
d\mathbf{s}
 \label{dfradiom}
\end{equation}
Here we have used the above mentioned approximation of an isothermal
TM and $\delta T(\mathbf{r})\ll T_{0}$. This simplified ``infinite
plate'' model assumes that all (and only) the molecules emitted by
the GRS surfaces directly in front of the TM (for the $X$ faces,
these are the red and yellow zones in \figr{tempgrad:ideal}) strike
the opposing TM face. As a consequence, edge effects are negligible,
and the net shear force acting parallel to any TM surface is zero.
Thus, the integrated radiometric contribution to the $x$ force comes
from the GRS surfaces on the $X_+$ and $X_-$ faces and is given by
\begin{eqnarray}
\label{Force:radiom} F_{Rx} & = & \frac{P}{4T_{0}}\sum_{X_{+},X_{-}}
\int_\mathrm{A}  \left( \mp \right) \delta T(y,z) dy dz
\nonumber \\
& = & \frac{\mathrm{A} P}{4}\left(\frac{\overline{T}_{X_{+}}
-\overline{T}_{X_{-}}}{T_0}\right)
\end{eqnarray}
where, as defined above, $\overline{T}_{X_{+}} -
\overline{T}_{X_{-}} = \Delta T_x$ is the average temperature
difference between the $X_{+}$ and $X_{-}$ GRS surfaces, and $A$ is
the the area of a TM face. For example, in the idealized temperature
difference distribution shown at left in \figr{tempgrad:ideal},
$\Delta T_x = 2 \delta T$.

The radiometric ``transfer function'' is thus
\begin{eqnarray}
\left . \frac{dF_x}{d({\Delta T}_x)} \right | & = & \kappa_R \frac{A
\, P}{4T_0}
\nonumber \\
& = & 18 \: \mathrm{pN / K} \times \kappa_R \left( \frac{P}{10^{-5}
\: \mathrm{Pa}} \right) \left( \frac{293 \: \mathrm{K}}{T_0} \right)
\label{radiom_numbers}
\end{eqnarray}
Here, we include a radiometric correction factor $\kappa_R$ in order
to incorporate modifications to this simple model that arise when
considering a realistic sensor geometry with finite dimensions,
which will be discussed shortly.  The factor $\kappa_R$ will be a
function of the sensor geometry, but also of the particular
temperature distribution inside the sensor that produces a given
average temperature difference $\Delta T_x$.

In this model, the torque component $N_z$, relevant to our torsion
pendulum experiment, is calculatd by integrating the moment of the
normal radiometric force element in \eqr{dfradiom} over the area of
the $x$ and $y$ TM faces, yielding
\begin{eqnarray}
\label{radiom:torqtheo} N_{z}\simeq \frac{P}{4T_{0}}& \cdot &\left(
\sum_{X_{+},X_{-}} \int_A  \left( \pm \right) \delta T(y,z) y dy dz
\right.
\nonumber \\
  && + \left. \sum_{Y_{+},Y_{-}}  \int_A  \left( \mp \right)
\delta T(x,z) x dy dz\right),
\end{eqnarray}
The sum of the integrals within the parentheses will be called
``thermal integral'' and indicated with $ \int_S T(\mathbf{s})
b(\mathbf{s})ds$. The arm-length $b(\mathbf{s})$ coincides, taking
the appropriate sign, with the coordinate $y$ on the $X$ faces and
the coordinate $x$ and on $Y$ faces. Thus \eqr{radiom:torqtheo} can
be simply written as an integral over the TM $X$ and $Y$ surfaces,
\begin{equation}
\label{radiom:torqtheosimp}
 N_{z}\simeq \gamma _R \frac{P}{4T_{0}}\int_S T(\mathbf{s})
b(\mathbf{s})ds.
\end{equation}
By analogy with the correction factor $\kappa_R$ introduced in
\eqr{radiom_numbers}, we include here the factor $\gamma_R$, as a
radiometric torque correction factor for a finite size sensor.
$\gamma_R$ will also be a function of the sensor geometry and of the
particular temperature distribution inside the sensor that produces
a given thermal integral.

To illustrate the significance of the ``thermal integral,'' we
consider the simple temperature distribution on the right in
\figr{tempgrad:ideal}, which is an idealization of the temperature
pattern imposed in our experiment. In this case,
\begin{equation}
\label{thermal:torquesimple} \int_S T(\mathbf{s})
b(\mathbf{s})ds=\delta T \frac{L^3}{2} =4\delta T \frac{L^2}{2}
\frac{L}{4} \, . \end{equation}
Each ``electrode'' (the half surfaces
of the $X$ faces) contributes with a torque proportional to the
temperature difference $\delta T$ multiplied by its area $L \times
\frac{L}{2}$ and the "arm" $\frac{L}{4}$ (that is the distance of
the center of the electrode from the center of the TM face). For the
same $\delta T$, the thermal integral scales as the cube of the
linear size of the TM.
\begin{figure}[t]
\includegraphics[width=45mm]{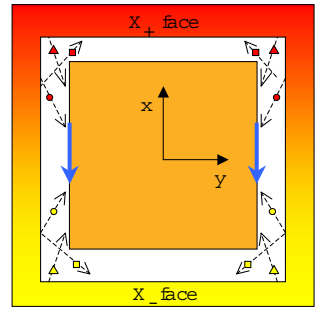}
\caption{\label{pictorials}Cartoon of the GRS cross section,
illustrating shear forces along $x$ caused by molecules striking the
$Y$ TM faces in the presence of a linear thermal gradient in $x$ (as
in the simulations of \aref{radiom:app}). Molecules striking the $Y$
faces impart shear forces parallel to the TM surface along $x$. For
a non-infinite TM, some molecules (such as those represented by
squares) emitted from a given point on $Y$ sensor surfaces do not
strike the TM. As such, this asymmetrizes the $x$ momentum imparted
to the TM by the molecules originating in the given sensor-face
location, which, in the presence of a thermal gradient produces a
net thermal shear force (blue arrows). This is reinforced by
molecules (represented by triangles) starting from the $X$ faces
near the corners. From the standpoint of a surface element of the
TM, there is a net imbalance in the shear momentum imparted by the
molecules arriving from the hot (red) and cold (yellow) sides that
contributes to the overall translational radiometric
effect. This shear effect increases with increasing gap size.\\
Note that the momentum components transferred normal and parallel to
the $Y$ surface contribute to the torque around the $z$ axis with
opposite sign.  As such, these shear forces are also responsible for
a reduction in the radiometric torque, relevant to the experiments
presented here.}

\end{figure}

For realistic sensor dimensions, finite size effects begin to erode
the infinite plate assumptions used here. First, not all molecules
leaving the GRS surface directly in front of the TM (for the $X$
faces, the yellow and red zones in \figr{tempgrad:ideal}) strike the
TM.  Other molecules originating in the corners (the grey corners in
\figr{tempgrad:ideal}) do strike the TM, on the $X$ faces but also
on the $Y$ and $Z$ faces. Additionally, molecules striking the $Y$
and $Z$ faces impart shear forces along $x$, which in the presence
of a temperature gradient is rendered asymmetric by the finite TM
dimension (illustrated in \figr{pictorials}). These phenomena are
enhanced with larger gaps and, in connection with
temperature-induced asymmetry of the molecular velocity
distribution, are expected to contribute to both $F_{Rx}$ and
$N_{Rz}$. To understand the role of these corrections to the simple
radiometric model presented above, we have performed a numerical gas
dynamics simulation, which is described in detail in
\aref{radiom:app}.

Simulations have been performed for a cubic 46~mm TM inside a
rectangular box, with several different gap sizes, including the
geometry for the LISA Pathfinder sensor design (shown in
\figr{sens}). The simulations consider a temperature difference
$\Delta T_{x}$ constituted by $X_+$ and $X_-$ faces that are at
constant temperature (differing by $\Delta T_{x}$), with a linear
temperature gradient along the $Y$ and $Z$ faces (as illustrated in
\figr{pictorials}). For the TM force along $x$, the simulations
indeed verify the simplified model of \eqr{radiom_numbers}
($\kappa_R = 1$), to within 10\%, for the smallest TM-GRS gaps
studied (1 mm, see \figr{simula:radiom} in \aref{radiom:app}), with
the force coming nearly exclusively from molecules impacting the $X$
faces. With increasing gap size, however, the calculation
demonstrates the role of shear forces along the $Y$ and $Z$ faces,
which contribute to an excess of roughly 40~\% above the infinite
plate model prediction for the largest gaps of the study (7~mm). For
the LISA Pathfinder GRS design, the radiometric force transfer
function $\left . \frac{dF_x}{d \left( \Delta T \right)} \right |
_R$ exceeds the infinite plate prediction \eqr{Force:radiom} by
roughly 25 \% ( $\kappa_R \approx 1.25)$, with a statistical error
of several percent (results are summarized in Table
\ref{table_alpha_cap}, along with those from a similar radiation
pressure simulation).

\begin{table}[h]
\begin{ruledtabular}
\begin{tabular}{ c  |  l  |  c c c }
$\kappa_R$      & & 1.25 \\
\hline
        & $a$ = 1 & 1.17 \\
$\kappa_{RP}$   & $a$ = 0.05 (specular) & 0.32 \\
                & $a$ = 0.05 (diffuse) &  0.75 \\
\end{tabular}
\end{ruledtabular}
\caption{\label{table_alpha_cap} Results of the thermal gradient
simulations, where $\kappa_{R}$ and $\kappa_{RD}$ represent
correction factors to the infinite plate model for, respectively,
the radiometric effect (\eqr{radiom_numbers}) and differential
radiation pressure (\eqr{rad_pres_numbers}). The simulations assume
the LTP GRS design geometry and a temperature difference between two
opposing GRS $X$ faces, each at a uniform temperature and with a
linear temperature gradient along the $Y$ and $Z$ faces. Results for
radiation pressure are shown for different values of the absorption
$a$ and nature of reflection (specular or diffuse). Statistical
errors are of order several percent.} \label{table_alpha}
\end{table}

The simulations employing a linear temperature gradient along $x$
also allow an estimate of the radiometric torque relevant to the
torsion pendulum experiment, as the linear temperature profile along
$y$ produces a torque around the $z$ axis. Here, a net decrease of
the radiometric effect is observed with respect to the simplified
model of \eqr{radiom:torqtheo} (see \figr{simula:radiom} of
\aref{radiom:app}). With increasing gap, there is a suppression of
the radiometric torque, dominated by the same shear forces mentioned
above, which tend to work against the torque contribution of the
dominant normal component of the molecular impacts (illustrated in
\figr{pictorials}). The magnitude of the torque reduction, with
respect to the infinite plate model, is in fact roughly consistent
with the shear force -relevant to the increase in radiometric force
reported above -multiplied by an arm length of $L$/2, or half the TM
width. For the $Y$ face torque contribution for the 2~mm gap sensor,
we find roughly 83\% of the torque calculated with the infinite
plate model (\eqr{radiom:torqtheo}), and only 65\% for the
larger-gapped LTP GRS prototype.

The true correction factor $\gamma_R$ will depend on the real
temperature profile for the experiment, rather than the contribution
of a single face that sees a linear temperature profile. However,
the experimental uncertainty in estimating the true temperature
profile in these experiments (see \sref{meas_tech} and
\aref{tempattern:app}), particularly in the corners of the sensor
where the finite-size corrections are most relevant, limits the
value of a simulation performed with a ``full'' (but very
approximate) sensor temperature distribution. A linear temperature
distribution also turns out to be a decent approximation for the
experimental temperature distributions. As such, the torque results
obtained for a linear temperature gradient, extrapolated to a full
sensor, yield indicative estimates, $\gamma_R \approx 0.65$ and
$\gamma_R \simeq 0.8$, for comparing the experimental results for
the torque measurements for, respectively, the LTP GRS and 2~mm
sensor prototypes.

\subsection{Thermal Radiation Pressure}
\label{rapress:theor}

The effect of thermal radiation pressure can be evaluated from the
momentum transfer of thermal photons emitted by all radiating
surfaces inside the GRS. We again assume an infinite parallel plate
approximation, in which all thermal photons wind up being absorbed
in the vicinity of emission, either on the parallel surface opposite
the point of emission or reflected back onto the original emitting
surface. Assuming the same emissivity for all the involved surfaces,
each surface element $d\mathbf{s}$ on the internal TM-facing
surfaces of the sensor contributes with a force
$dF_{RP}(\mathbf{r})$ acting on the TM in the direction
perpendicular to the surface element itself. This force is given by
\begin{eqnarray}\label{radiation:2}
 dF_{RP}(\mathbf{r})=\frac{2\sigma}{3c} T^4(\mathbf{r})d\mathbf{s}
\simeq  \frac{2\sigma}{3c} T_{0}^4(\mathbf{r})d\mathbf{s}+
\frac{8\sigma}{3c} T^3_0 \delta T(\mathbf{r})d\mathbf{s}
\end{eqnarray}
with $\sigma$ the Stefan-Boltzmann constant and $c$ the speed of
light. Assuming that the TM is isothermal, and thus contributes no
net recoil force from its radiation, the net force acting on the TM
in the sensitive $x$ direction for LISA will be proportional to the
average temperature difference between opposing sensor faces, given
by
\begin{equation}\label{radiation:press}
    F_{RP_x}
\simeq \mathrm{A} \frac{8\sigma}{3c}T^3_{0}
\left(\overline{T}_{X_{+}} -\overline{T}_{X_{-}}\right).
\end{equation}
where \eqr{radiation:2} is integrated over the EH surfaces that
directly face the TM (the yellow and red zones in the left image in
\figr{tempgrad:ideal})\footnote[1]{\eqr{radiation:press} is
independent of the absorptivity coefficient if the TM and sensor
wall absorptivities, $a_{TM}$ and $a_{S}$ respectively, are equal.
In the event that they are not equal, \eqr{radiation:press} is
modified by a factor $\frac{a_{S} \left(2 - a_{TM} \right)} {a_{TM}
+ a_{S}- a_{TM} a_{S}}$, which can range from 0 to 2.}. Thus, the
radiation pressure transfer function can be expressed
\begin{eqnarray}
\label{rad_pres_numbers}
 \left . \frac{dF_x}{d \left( {\Delta T}_x \right)} \right| _{RP}
& \simeq & \kappa_{RP} \mathrm{A} \frac{8\sigma}{3c}T^3_{0} \nonumber \\
& \simeq & 27 \: \mathrm{pN/K} \times \kappa_{RP} \left(
\frac{T_0}{293 \: \mathrm{K}} \right)^3 \label{fresp_rp2}.
\end{eqnarray}

As in \sref{radiom:theor}, we can integrate the moment of the force
element $dF_{RP}$ (\eqr{radiation:2}) to calculate the torque $N_z$,
\begin{eqnarray}
\label{radiation:3} N_{RPz}\simeq \gamma_{RP} \frac{8\sigma}{3c}
T_0^3 \int_S T(\mathbf{s}) b(\mathbf{s})ds.
\end{eqnarray}
where both the $X$ and $Y$ faces contribute.  We note that in
analogy with the radiometric effect in the previous section, we have
introduced, in Eqns. \ref{rad_pres_numbers} and \ref{radiation:3},
respectively, we have introduced the radiation pressure force and
torque factors $\kappa_{RP}$ and $\gamma_{RP}$, to account for
corrections due to the inaccuracy in applying this simple infinite
plate model to our true sensor geometry.

As with the radiometric effect, the assumptions for this simple
radiation pressure model begin to break down when the gap sizes are
not negligible compared to the TM dimensions.  Just as for gas
molecules, there are corrections due to emitted photons ``missing''
the TM and hitting adjacent faces, as well as adsorbed photons along
the lateral ($Y$ and $Z$) faces imparting a shear force and
associated torque on the TM.  In addition, the presence of finite
reflectivity (absorptivity $a < 1 $ means that bouncing thermal
photons will hit multiple surfaces, imparting force and torque
components with different signs, before adsorbing.  This latter
effect is particularly significant for the metallic surfaces inside
the sensor, which are expected to have a high 90-95 \%
reflectivities ($a \approx 0.05 - 0.1$) at thermal photon
wavelengths \cite{radiaprop}, and systematically serves to
homogenize the radiation pressure felt by the TM, suppressing the
differential radiation pressure effect.

To properly analyze such effects, a numerical simulation was
performed and is described in detail in \sref{rapress:app}. The
simulation studied several representative simplified sensor
geometries and sensor temperature profiles representative for the
force for LISA and for the torque relevant to the experiments
presented in the following sections. Additionally, different extreme
cases for the infrared absorption $a$ and for the nature of the
reflection (specular or diffuse) were considered. For the force, a
simple linear temperature gradient along the $x$ axis was assumed,
with uniform temperature across each $X$ face. In the limit of
vanishing TM-sensor gaps (0.2 mm), the infinite parallel plate model
for the force (\eqr{radiation:press}) was confirmed, with the factor
$\frac{8\sigma}{3 c}$ (or $\kappa_{RP} \approx 1$), to within
several percent, even for absorption as small as 0.1. For perfect
absorption ($a$ = 1), there is an increase in the radiation pressure
effect akin to that for the radiometric effect, with $\kappa_{RP} =
1.17$ for the LISA Pathfinder geometry (see the summary of results
in \tabr{table_alpha}). For the case of the high reflectivity
surfaces expected for the LISA sensor -- using $a = 0.05$, with
specular reflection -- there is a suppression of the effect, with
$\kappa_{RP} = 0.32$. This result thus indicates a significant, if
not order-of-magnitude, change to the estimate of the effect given
in simplified radiation pressure models for LISA
\cite{stebbins:errors,bonnie:noise,dolesi:sensor}.

For the torque, assuming temperature profiles similar to that
estimated in our experiments (using the technique described in
Appendix A), the suppression of the differential radiation pressure
effect is more pronounced, with photons moving, in only a couple
bounces, from one half of the TM to the other (or to an adjacent
face), inverting the sign of the resulting torque. For the high
reflectivity surfaces ($a = 0.05$ and specular reflection)
reasonable for the gold coated TM and sensors used, we obtain
radiation pressure torque correction factors $\gamma_{RP}$ of order
0.4 and 0.3 for the 2 mm and 4 mm (LISA Pathfinder design)
prototypes used in the experiment. This significant reduction has an
impact on the interpretation of the experimental results discussed
in \sref{exp_results_disc}.

\subsection{Asymmetric temperature dependent outgassing}
\label{asy_outgassing} Outgassing of molecules absorbed by the
internal walls of the GRS increases the residual pressure
surrounding the TM. Additionally, an asymmetry in the molecular
outflow can be created or enhanced by the temperature dependence of
the outgassing, resulting in a differential pressure that is driven
by the GRS temperature gradient itself.

As asymmetric outgassing depends on the amount and type of absorbed
impurities inside the GRS, it is a contamination effect that does
not lend itself easily to a first principles calculation, in
contrast with the previous two effects.  However, the gas flow from
a surface can be modelled with a temperature activation law
$\mathcal{Q}(T)=\mathcal{Q}_0 e^{(-\Theta/T)}$, where
$\mathcal{Q}_0$ is a flow prefactor and $\Theta$ the activation
temperature of the molecular species under consideration
\cite{rudiger,vitale:therm:simu}. Our simplified model considers
only a single outgassing species which is uniformly desorbed from
identical GRS surfaces, and that the molecules emitted from the
sensor walls collide with the TM without sticking. A temperature
gradient $\Delta T_x$ would induce an asymmetric rate of outgassing
across the TM $\Delta \mathcal{Q}(T)\approx
\mathcal{Q}(T_0)\left(\Theta /T_0\right) \left(\Delta T_x/T_0
\right)$. The consequent difference in pressure $\Delta P_{out}$
across the TM would cause a force $F_{OG x}$ given by
\begin{equation}\label{outgas:press}
  F_{OG x}=A\,\Delta P_{OG}=A \frac{\Delta
  \mathcal{Q}(T)}{C_{eff}}\approx A
\frac{\mathcal{Q}(T_0)}{C_{eff}}\frac{\Theta}{ T_0} \frac{\Delta
T_x}{T_0}
\end{equation}
where $C_{eff}$ is a geometrical factor resulting from a combination
of the conductance of the paths around the TM and through the holes
in the GRS electrode housing walls, estimated to be roughly 4.3
$\times 10^{-2}$ m$^3$/s for the LISA / LISA Pathfinder GRS geometry
\cite{dolesi:sensor}. The asymmetric outgassing transfer function is
then given by
\begin{eqnarray}
\left. \frac{dF_x}{d \Delta T}_x  \right|_{OG}& \simeq & A
    \frac{\mathcal{Q}(T_0)}{C_{eff}}\frac{\Theta}{ T_0^2} \nonumber \\
    & \simeq & 40 \, \frac{pN}{K}
\left( \frac{\mathcal{Q} (T_0) }{1.4 \, \mathrm{nJ/s}} \right)
\left( \frac{\Theta}{3 \times 10^4 \, \mathrm{K}} \right) \left(
\frac{293 \, \mathrm{K}}{T_0} \right)^2
\end{eqnarray}
We note that the temperature dependence of the effect, partially
hidden in $Q(T_0)$, is $\frac{1}{T^2} \exp{\left( -\Theta/T \right)
}$, at least in this simplified model with a single outgassing
species, and is thus a rapidly increasing function at the
temperatures of our study. The values used for the flow
$\mathcal{Q}(T_0)$ and the activation temperature $\Theta$
correspond to typical room temperature behaviors for the metal and
ceramic materials used in the GRS \cite{Roth:book, hanlon:book}.
This source of coupling to the thermal gradient thus appears to be
roughly equal in magnitude to that of the combined radiometric and
radiation pressure effects (Eqns. \ref{radiom_numbers} and
\ref{rad_pres_numbers}). However, given the likely large variability
due to materials, assembly, cleaning, and thermal history of the
sensor, this noise source has a much larger uncertainty and is a key
motivation for measurement of the overall thermal gradient transfer
function. If the outgassing phenomenon were indeed uniform across
the GRS surfaces, the torque transfer coefficient could be found by
integrating the moment due to the differential outgassing pressure
over the sensor surfaces, as for the two preceding thermal gradient
effects. This would allow conversion between force and torque
through an effective arm length, of order $\frac{L}{4}$ for a
simplified temperature distribution like that in
\figr{tempgrad:ideal}. However, unlike the previous two effects, the
numerous features in the sensor -- holes, interfaces between the
metal and ceramic surfaces -- suggest that the outgassing could be
concentrated in specific localized areas, rather than uniform. As
such, there is no simple, unequivocal extrapolation of
 $\left. \frac{dF_{x}}{d \Delta T_x} \right|_{OG}$ from a torque measurement. However,
the torque measurements are sensitive to nearly all possible
outgassing effects, regardless of their model, and the lack of a
significant excess beyond the well-modelled radiometric and
radiation pressure effects will certainly build confidence that
there is no dominant asymmetric outgassing effect for LISA.

\section{Experimental Details}\label{experiment}
\subsection{Gravitational Reference Sensor prototypes}\label{GRS:prototypes}

This experimental campaign studied two GRS prototypes realized as
part of the sensor design development for LISA. They feature
different electrodes geometries, construction techniques, and TM
sizes. Both sensors are based on a Gold coated Molybdenum housing,
with insulating elements made from the Aluminum-based, high thermal
conductivity ceramic SHAPAL. The first prototype sensor, referred to
here as ``TN2mm'' uses a 40mm cubic TM, with a gap -- the distance
between the TM and the surrounding electrode and electrode housing
surfaces -- of 2~mm, symmetric on all three axes. The electrodes are
Au-coated bulk Mo plates separated from the grounded housing by bulk
Shapal plates. The second prototype, ``EM4mm,'' reflects the final
GRS design for the LTP mission, with a 46mm cubic TM and gaps of 4,
2.9 and 3.5 ~mm on, respectively, the $x$, $y$, and $z$ axes (see
\figr{sens}). In contrast with TN2mm, the electrodes for EM4mm are
obtained by Au sputtering deposition directly on the bulk insulating
SHAPAL plates.

\subsubsection{Electrical heaters and thermometers}
\label{GRS:heatandterm}
\begin{figure}
\includegraphics[width=80mm]{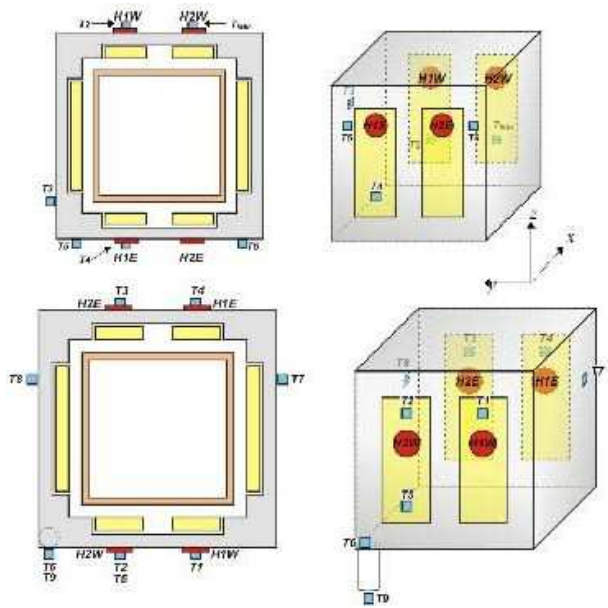}
\caption{\label{therm:loc}Top panel: prototype TN2mm was equipped
with five thermometers and the experiment was performed by
alternatively switching on and off the heaters H1E and H1W (red
circles). Bottom panel: prototype EM4mm, equipped with eight
thermometers. For prototype EM4mm we employed all four heaters,
alternatively switching the pairs (H1E,H2W) and (H2E,H1W). We note
that $T_{fake}$ is included here on the TN2mm $X+$ face to indicate
the position of a temperature reading that, as described in
\aref{tempattern:app}, has been artificially added to help estimate
the temperature profile, in the absence of a thermometer near to
heater H2W. }

\end{figure}

Electrical heaters have been installed on the external surfaces of
both sensor prototypes, in correspondence with the $X$ face
electrodes.  They are displaced with respect to the pendulum torsion
axis (see \figr{therm:loc}) and can thus apply the rotational
temperature differences that convert the forces described in
\sref{estim:theor} into torques, to which this pendulum is
sensitive. The heaters are made from of twisted pairs of Manganin
wires several meters long, wound onto 5~mm diameter Al cylinders.
The heaters are attached to the sensor with a high thermal
conductivity, vacuum compatible glue, for the TN2mm prototype, and
tightly screwed onto the GRS surfaces for EM4mm. The resistance of
each heater is $95 \pm 1$ Ohms. A 1~Hz square wave voltage, up to a
maximum of 7~V across the resistor (nearly 0.5~W), is supplied to
the heaters with high output current operational amplifiers.
Twisting the heater wires limits the magnetic field produced by the
heater current, and the use of the 1~Hz square wave allows
application of constant heater power, while at the same time
converting any residual magnetic field-related torque to a frequency
well above our measurement frequency of 0.4-0.5~mHz (see
\sref{meas_tech}).

With the aim of reconstructing the temperature profile of the GRS,
temperatures at different positions are measured for each sensor, by
a set of Pt100 thermometers on the electrode housing external
surfaces. They are glued to the sensor by means of a high thermal
conductivity vacuum compatible glue. Five thermometers were used on
the TN2mm prototype and eight on EM4mm. A sketch of the thermometer
and heater locations is shown for both sensors in \figr{therm:loc}.

\subsection{Torsion pendulum apparatus}

\begin{figure}
\includegraphics[width=85mm]{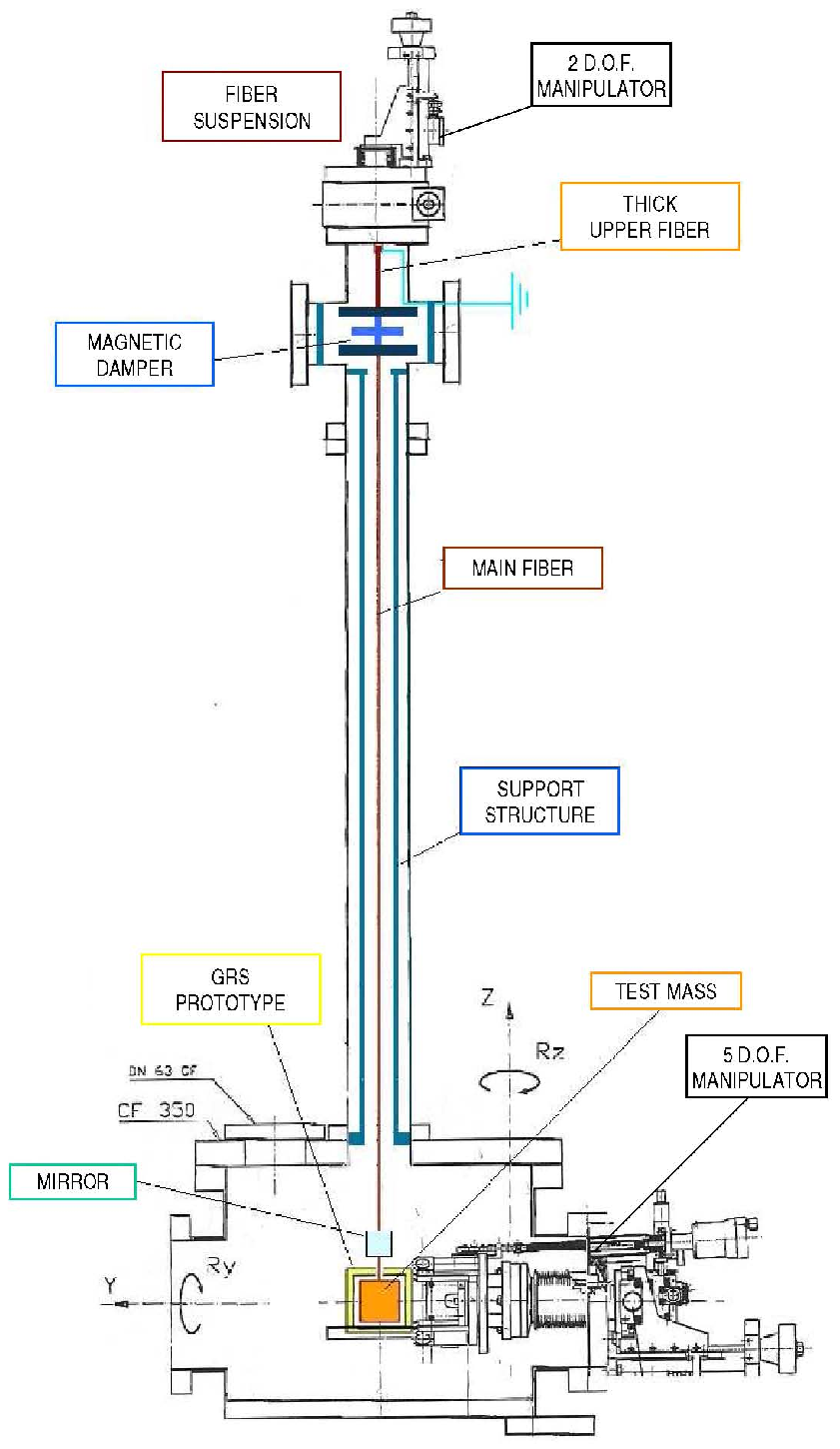}
 \caption{\label{facility} Schematic of the torsion pendulum apparatus.
The TM is the main part of the inertial member of a torsion pendulum
hanging from a long torsion fiber designed to have the minimum
stiffness and thus the maximum sensitivity. The electrode housing
surrounds the TM and is attached to a precision positioner that can
center it relative to the TM in 6~degrees of freedom.}
\end{figure}
This experimental investigation has been performed with a torsion
pendulum facility
\cite{hueller:pend,prd:force_noise,carbone:prl,hueller:upperlimits,carbone:thesis,carbone:char},
where a representative copy of the LISA test mass is suspended by a
thin fiber and hangs inside a prototype of the GRS. The small
torsional elastic constant of the fiber allows for high resolution
measurement of torques around the fiber axis, via measurement of the
pendulum angular deflection. Measurements of the pendulum angular
deflection noise can be used to study the force noise acting on the
LISA TM. Additionally, controlledly modulation of known disturbance
fields, such as the thermal gradients in this study, with coherent
detection of the resulting pendulum deflection, can allow a study of
known force noise sources relevant to LISA free-fall.

The pendulum is designed for minimum elastic stiffness and thus
maximum torque sensitivity. A light-weight hollow TM (Au-coated Ti
for TN2mm, and Au-coated Al for EM4mm) has been chosen to maximize
the torque sensitivity, allowing use of a thin (25~$\mu$m) tungsten
fiber, with a resulting resonant frequency near 2~mHz and quality
factor near 3000. The pendulum and sensor are mounted on independent
micro-positioners, allowing a 6~degree of freedom adjustment of the
relative position of TM inside the electrode housing. The torsion
pendulum deflection angle was read out, with 10~Hz sampling rate, by
an optical autocollimator measurement of a mirror that was part of
the torsion member.

The vacuum chamber that accomodates the pendulum and the GRS is
evacuated by means of a turbomolecular pump, and its pressure is
monitored by means of an ionization gauge (Varian UHV 24). The whole
apparatus sits in a thermally controlled room with 50 mK long term
stability.

The pendulum torque resolution relevant to these measurements is
typically 5-10~fN~m~/~Hz$^{1/2}$ near the 0.4 - 0.5~mHz heater power
modulation frequency, which is a factor 2-3 above the Brownian
thermal noise limit for the pendulum. For the several hour
integration times used in these measurements, this equates to a
torque resolution of order 0.1~fN~m.

\begin{figure}[t]
\centering
\includegraphics[width=80mm]{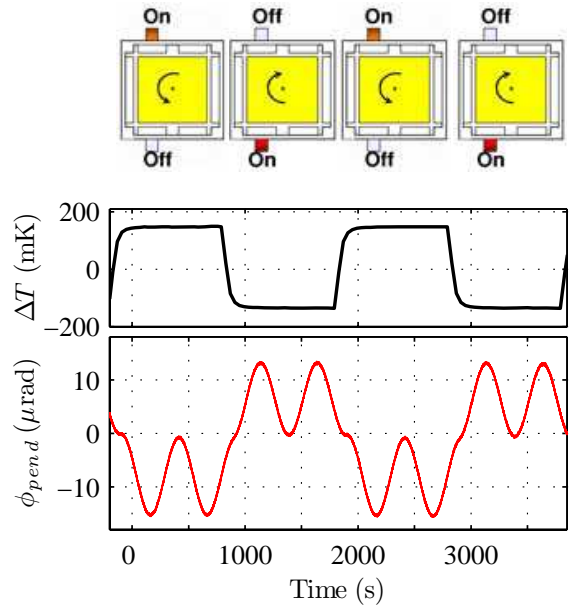}
\caption{\label{fig1} Illustration of the thermal gradient effects
measurement technique for the the TN2mm prototype. Here 200~mW are
alternatively applied to heaters (H1E and H1W in \figr{therm:loc})
located off-axis on opposing sensor faces, switching at 1000~s
intervals (top panel). The central panel shows the induced
temperature difference $\Delta T = \left ( T_2-T_4 \right ) $. The
resulting thermally induced torque causes the pendulum to respond
coherently, as shown in bottom panel, with an angular deflection at
the heater modulation frequency superimposed on the pendulum free
oscillation.}
\end{figure}

\subsection{Measurement technique and data analysis}
\label{meas_tech}

The general experimental technique, illustrated in \figr{therm:loc},
is to create an oscillating asymmetric temperature distribution by
modulated heater powers and then measure the resulting coherent
oscillations both in the torque on the TM and in the GRS temperature
distribution. Varying and monitoring the average sensor temperature
and pressure, in addition to the oscillating temperature difference
pattern, allows separation of the different thermally induced torque
effects for their dependence on these parameters. In presenting the
measured torque data, they will be normalized for the amplitude of
the applied temperature gradients as expressed by the estimated
applied ``thermal integral,'' the integral $\int_S T(\mathbf{s})
b(\mathbf{s})ds$ introduced in \sref{estim:theor} to characterize
the radiometric and radiation pressure torques in the infinite plate
limit.  This normalization allows comparison between measurements of
different thermal gradient amplitudes and comparison with the models
for the radiometric and radiation pressure effects discussed in
\sref{estim:theor}.

\subsubsection{Rotational thermal gradients: application and estimation}
\label{tech:DeltaT}

For the TN2mm prototype, the oscillating temperature pattern were
obtained by alternately switching on and off two opposing heaters
for 1000~s, as illustrated in \figr{therm:loc}. For the EM4mm
prototype, two pairs of opposing heaters are switched on and off for
1250~s. These modulation frequencies, 0.5 and 0.4~mHz respectively,
result in coherent pendulum torques that are below the roughly 2~mHz
resonant frequency. Typical peak to peak temperature differences
across the sensor ranged from tens to hundreds of mK. The alternate
switching of the heaters produce an oscillating temperature gradient
pattern across the EH, with the time dependence of a lightly
filtered square-wave. We note that the thermal time constant for
equilibration across the GRS, relevant to the temperature difference
(as plotted in \figr{therm:loc}) and to the thermal integral, is of
order 100~s.

The temperature pattern was recorded in the readings $T_i(t)$ of
each thermometer $i$ at its position $(s_{1},s_{2})_i$, where $s_1$
and $s_2$ are the two coordinates mapping the 4 $X$ and $Y$ GRS
surfaces. For each cycle $k$ of the heater power modulation, we
calculate the component $T(s_{1},s_{2},\omega,k)_i$ of $T_i(t)$ at
the modulation carrier frequency and at its third and fifth
harmonics, where the filtered square wave thermal pattern is known
to have significant spectral content. An estimation of the
components $T(s_{1},s_{2},\omega,k)$ at a generic position $(s1,s2)$
was then obtained by spatial interpolation of the values
$T(s_{1},s_{2},\omega,k)_i$ (for details see the appendix
\aref{tempattern:app}). For each demodulation cycle $k$, it was then
possible to obtain an estimation of the $\omega$ component of the
``thermal integral'', as defined in \eqr{radiom:torqtheo} and
\begin{eqnarray}
  \label{T_thermal}
\left(\int_S T(s) b(s) ds \right)_{(\omega,k)}& \, &  =
\sum_{X_{+},X_{-}}\nonumber \int_A (\pm)  T(y,z,\omega,k) y dy dz
+\nonumber\\
& + &  \sum_{Y_{+},Y_{-}} \int_A \left( \mp \right) T(x,z,\omega,k)
x dy dz
\end{eqnarray}
As discussed in appendix \aref{tempattern:app}, our analysis shows
that the simple interpolation technique used underestimates the
thermal integral, by 20-30~\% in the simplified analysis used.
However, given the coarse sampling of the temperature distribution
with only 5-8 thermometers and the approximate nature of the
analysis used, we have chosen not to ``correct'' the data presented
for this systematic error, but rather to present the data obtained
from the relatively straightforward interpolation. The uncertainties
in the estimation of the thermal integral are of order 20\% and 50\%
for, respectively, the EM4mm and TN2mm sensors. Given the repeated
use of the same heating pattern for each of the two sensor datasets,
the uncertainty in the thermal integral estimation can be thought of
a single scale factor for each dataset.

\subsubsection{Torque estimation}
\label{tech:torque} The thermal gradient induced torque on the TM is
detected in the coherent deflection of the pendulum. The torque
components are calculated for each cycle $k$, at the odd harmonics
of the heater power modulation frequency, using the torsion pendulum
transfer function, as in Refs. \cite{carbone:prl} and
\cite{carbone:char}. The resolution of order 0.1fN m obtained for a
several hour integration corresponds to a signal-to-noise ratio of
order 10 for the smallest thermal gradient-induced torques measured.

The torque data for each cycle are normalized by division by the
calculated thermal integrals,
\begin{equation}
\label{torque_norm}
 N^{*}_{(\omega ,k)}=\frac{N_{(\omega ,k)}}{\left(
\int_S T(s) b(s) ds  \right) _{(\omega,k)}}
\end{equation}
The units of $N^{*}_{(\omega ,k)}$ are $\frac{Pa}{K}$.

\subsubsection{Average temperature: control and estimation}
\label{tech:aveT} Average temperatures between 283 K and 314 K were
obtained by changing the temperature of the room containing the
pendulum apparatus and by changing the total DC power applied to the
heaters. The average temperature of the electrode housing and TM
$T_0$ is derived by averaging the thermometer data for each
measurement cycle. In particular, average temperature data for
UTN2mm were obtained with $T_{0}=\frac{T_2+T_4}{2}$, whereas for
EM4mm we used $T_{0}=\frac{T_1+T_2+T_3+T_4}{4}$ (see
\figr{therm:loc}.

\subsubsection{Average pressure: control and estimation}
\label{pressure} The pressure in the vacuum enclosure were set to
different equilibrium levels, from roughly $2 \times 10^{-6}$ Pa to
$10^{-4}$ Pa, by changing the conductance between the vacuum vessel
and the pumping system. The pressure in the vacuum vessel was
measured by an ionization gauge (Varian UHV 24), sampled at 1 Hz and
averaged for each cycle of the heater modulation. The calibration of
the ion gauge, for which no certification was available, was
obtained by testing against a similar device (Pfeiffer UHV), that
was certified by the manufacturer to within 15$\%$. It was not
possible to use the Pfeiffer gauge constantly because of its very
high filament current, which produced an intolerable electrostatic
charging of the TM. The calibration was performed by simultaneously
acquiring both gauges while slowly changing the pressure in the
vacuum vessel with a valve on the pumping line, over a pressure
range from roughly $5\times10^{-6}$ Pa up to $2\times 10^{-4}$ Pa,
roughly the same range spanned by our study. The resulting fit of
the obtained curve was applied to calibrate the Varian gauge in all
measurements.

\section{Experimental Results and Discussion}
\label{exp_results_disc} 
\begin{figure}[t]
\centering
\includegraphics[width=85mm]{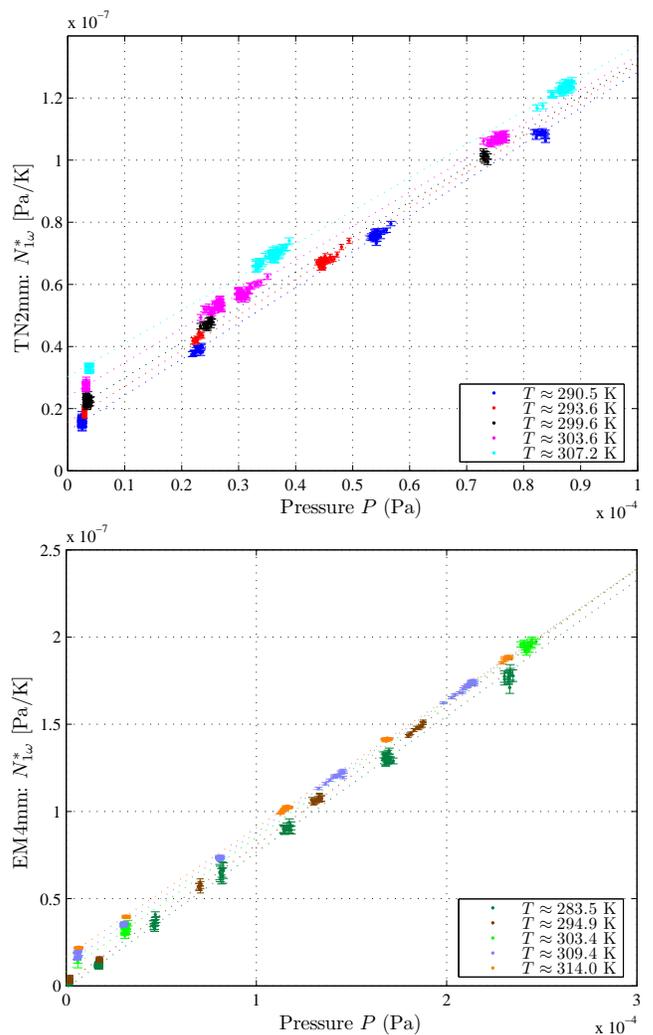}
\caption{\label{all_data}Normalized torque $N^{*}_{(\omega ,k)}$ of
\eqr{torque_norm}, as a function of pressure at different average
temperatures. Top panel the data for TN2mm prototype sensor, the
bottom panel the data for GRS Engineering Model for the LTP ,
indicated with EM4mm. Dashed lines are the results of linear least
square fits to the pressure at each average temperature. Only data
for the first harmonic of the excitation frequency are shown. }
\end{figure}

Results of the experimental investigation
for both the TN2mm and EM4mm sensors are shown in \figr{all_data},
where we report $N^{*}_{(\omega ,k)}$, the component of the
normalized torque at the heater modulation carrier frequency
$\omega$, as a function of pressure and for several average sensor
temperatures. The data are dominated by the linear pressure
dependence of the radiometric effect, which will be discussed in the
following paragraphs.

\begin{figure}
\includegraphics[width=85mm]{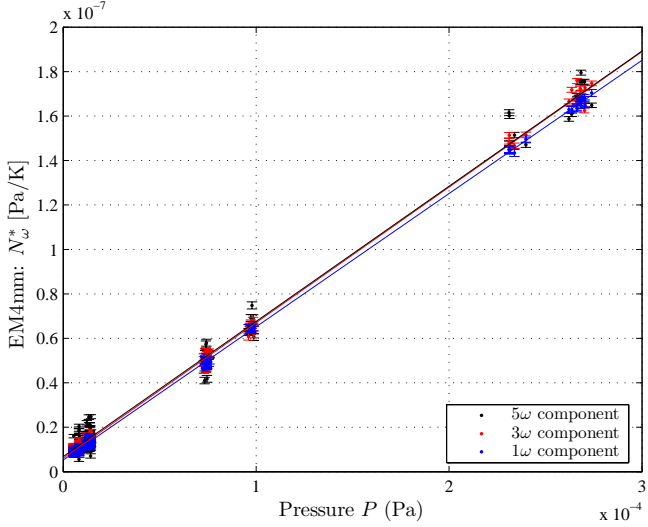}
\caption{\label{freq:compare} Normalized torques $N^*_{\omega,k}$ as
function of the residual pressure $P$, at the average temperature
$T_{0}\approx 298.9$K, calculated at the fundamental frequency
$1\omega=0.4$mHz (blue dots) and its odd harmonics $3\omega=1.2$mHz
(red) and $5\omega=2$mHz (black) (data for the EM4mm sensor). Linear
fits also are shown for comparison. The three measured slopes and
the three intercepts agree with each other within their statistical
errors (smaller than $4\%$).}
\end{figure}

\begin{figure}[t]
\centering
\includegraphics[width=85mm]{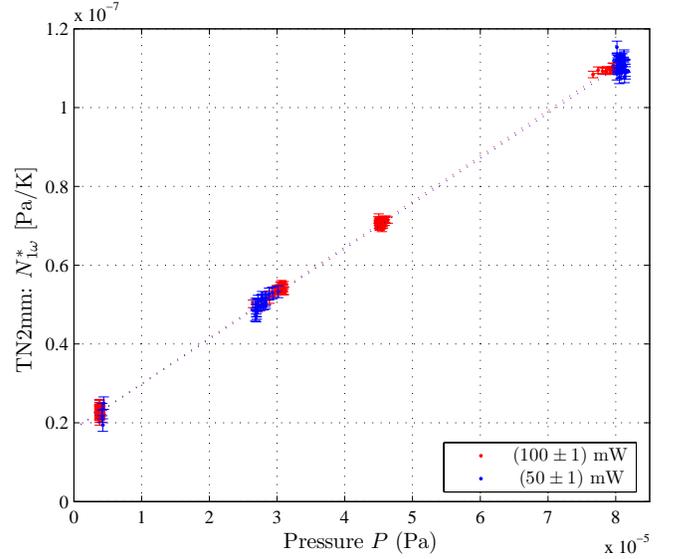}
\caption{\label{DT:diff} Measurement of $N^{*}_{(\omega ,k)}$ as a
function of pressure, with prototype TN2mm, for different values of
applied heat. Red points: alternating $200\pm 2$~mW between heaters
H1E and H1W. Blue points: apply $50\pm1$~mW continuously to both
heaters and alternating $100\pm 1$~mW. Lines are best fits to data.}
\end{figure}

In \figr{freq:compare} we plot the normalized torque data at the
first, third, and fifth harmonics of the modulation frequency (0.4,
1.2 and 2.0 mHz) -- each normalized for the thermal integral at the
relevant frequency -- with the 298.9~K data for the EM4mm sensor
chosen as an example. The data at the three frequencies are
equivalent within the statistical errors of the measurement. In this
study, no frequency dependence in the transfer function between
measured thermal gradients and resulting torque has been observed.
Also, by varying the amplitude of the modulated heater power, we can
check the linearity of the thermal induced torque effects in the
applied thermal integral amplitudes. One example is shown in
\figr{DT:diff} for the prototype TN2mm, showing that the normalized
torque data are reproduced, within the statistical errors, for the
two different temperature gradient amplitudes. These two
observations confirm the linear, frequency-independent nature of the
thermal gradient torque models discussed in \sref{estim:theor}.

In the following two sections we discuss the experimental results in
light of the models for the radiometric, radiation pressure, and
temperature-dependent outgassing effects discussed in
\sref{estim:theor}. From Eqns. \ref{radiom:torqtheosimp} and
\ref{radiation:3}, we expect that
\begin{equation}
\label{tottorque_norma} N^{*}_{(\omega ,k)} \simeq \gamma_R
\frac{P}{4 T_0}+\gamma_{RP}\frac{8\sigma}{3c} T_0^3,
\end{equation}
with an additional pressure independent term coming from
temperature-dependent outgassing. By fitting the data for $N^{*}
\left(T_0,P \right)$ in \figr{all_data} to the linear pressure
dependence at each average GRS temperature $T_0$, we first study the
radiometric effect, through the pressure slope. Finally through the
zero-pressure intercept, we can study the radiation pressure effect
and any outgassing related phenomena.

\subsection{Comparison with the radiometer effect model}
According to \eqr{tottorque_norma}, for the radiometer effect
$N^{*}(\omega ,k)$ should depend linearly on pressure with a slope
given by
\begin{equation}
\label{slope_radiom} \left ( \frac{\partial N^*_{\omega}}{\partial
P} \right)_{T_{0}} = \frac{\gamma_R}{4 T_{0}}
\end{equation}
where the correction factor $\gamma_R=1$ in the limit of the
infinite plate model. \figr{all_data} does indeed show that
$N^{*}(\omega ,k)$ depends linearly on $P$, with a slope that
decreases with $T_{0}$ (barely visible by eye in the plotted data).

\begin{figure}
\includegraphics[width=80mm]{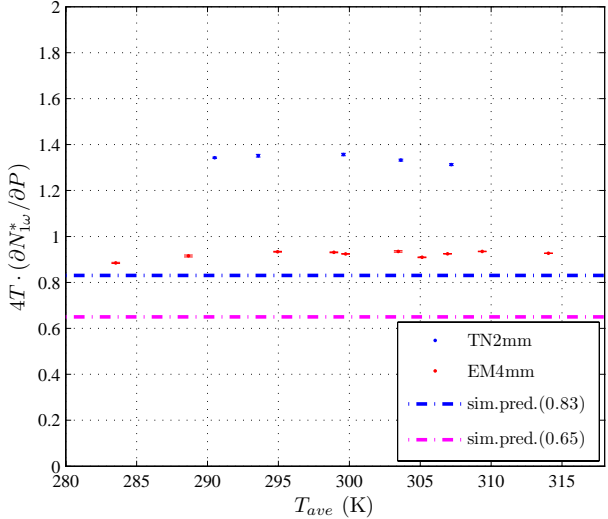}
\caption{Normalized measured slopes $4 T_{0} \frac{\partial
N^*_{\omega}}{\partial P}$ of the fitting lines in \figr{all_data}
as a function of the average temperatures $T_{0}$ (blue dots:
prototype TN2mm; red dots: prototype EM4mm). The results represent
measurements of the radiometric correction factor $\gamma_R$, which
is unity in the infinite plate model. Horizontal dashed lines:
values predicted by the simulation for a linear thermal gradient
(blue and red refer to TN2mm and EM4mm, respectively) from
\aref{radiom:app}.  We note that the error bars represent only the
statistical uncertainties in the torque measurement. The systematic
uncertainty in the estimate of the relevant thermal integral is much
larger, of order 20\% and 50\% for the EM4mm and TN2mm prototypes,
and serves as an uncertainty in the overall scale factor for each of
the two datasets.} \label{radiom}
\end{figure}

In \figr{radiom}, we plot the slope $\left ( \frac{\partial
N^*_{1\omega}}{\partial P} \right)_{T_{0}}$ multiplied by $4\,
T_{0}$, which thus represents the measured $\gamma_R$, a constant
according to our model. The results yield $\gamma_R = 1.34$ for the
TN2mm prototype and $\gamma_R = 0.92$ for EM4mm. The RMS fit
residuals are of order 1.5\% for each sensor, resulting in overall
uncertainties in $\gamma_R$ slightly below 1\%. We note that, for
both sensors, the observed scatter in these data exceeds that
estimated by the statistical torque uncertainty, by a factor 4-8. As
such these data are not limited by the torque resolution, but likely
by systematic uncertainties in the temperature integral and pressure
measurements between points at different temperature. We note that
the temperature ranges for both sets of data are sufficient to
distinguish the temperature dependence, and the fit to the
$\frac{1}{T_{0}}$ model (\eqr{slope_radiom}) is significantly better
than a fit to a constant (temperature independent) or to
$\frac{1}{T_{0}^2}$.

Figure \ref{radiom} also presents the values for $\gamma_R$
calculated in the radiometric simulations for the two sensor
geometries, assuming a simplified linear temperature gradient
(\sref{radiom:app}), which yielded $\gamma_R = 0.83$ for TN2mm and
$\gamma_R = 0.65$ for EM4mm. In comparing the experimentally
measured values to these predictions, we have to consider the
systematic uncertainties in the estimate of the thermal integral
that has been used to normalize the torque data.  The temperature
interpolation scheme used to estimate the thermal integral,
discussed in Appendix C, is likely to underestimate the thermal
integral. By one approximate analysis, this underestimate could be
roughly $20-30\%$ for the two sensors. Underestimating the thermal
integral would systematically increase the values of $\gamma_R$
obtained here, effectively scaling each of the two sensor data
curves shown here (as each data set relied on a single pattern of
heating and thermometer sampling, it is likely that each data set
would have a systematic error characterized by a single scale
factor).  Without a more reliable model for calculating the thermal
integral from the limited number of thermometers on the sensor, the
data presented use the simple temperature interpolation, with an
estimated systematic uncertainty of order $50\%$ and $20\%$ for the
TN2mm and EM4mm sensors, respectively.
 As such,
the data for both sensors are consistent with the predictions for
the radiometric effect. The uncertainties are however too large to
test the radiometric model at a level approaching the statistical
resolution of the measurements,  or to test definitively the
modifications to the infinite plate model predicted by the
simulations for the two sensors.

In addition to confirming the model of the radiometric effect, these
data confirm, and extend -- to a wider range of pressure,
temperature, and to a second prototype sensor -- our preliminary
observations with the TN2mm sensor
\cite{carbone:thesis,carbone:char}. We note that a thermal
gradient-induced torque that increases with pressure is also
observed in a LISA-related study at the University of Washington
(UW) \cite{pollack:thermal}, for measurements at two pressures and
two temperatures. This study, which employed a much simpler and more
open geometry than the LISA GRS, observed forces of the same order
of magnitude as those observed here. However the authors report an
unexpected frequency dependence, which is not observed at all in the
measurements presented here. In addition, the UW study reports an
increase in the pressure dependent effect with increasing
temperature, which is contrary to the expected radiometric
$\frac{1}{T}$ dependence observed here. Given the differences in the
experimental geometry and the limited number of presented
measurements, it is not possible at this time to further discuss
these apparent discrepancies.

\begin{figure}
\includegraphics[width=85mm]{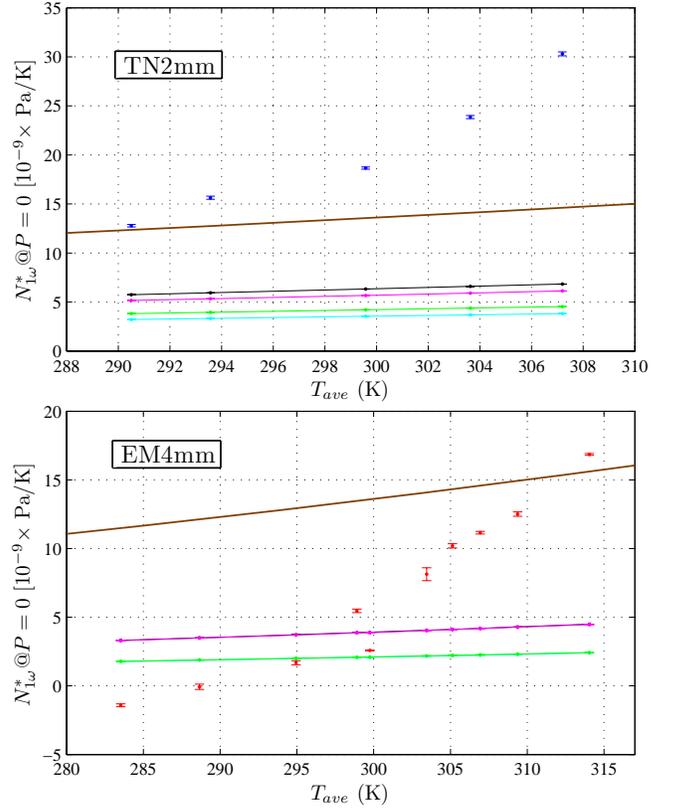}
\caption{\label{interc} Measured intercepts of the linear fittings
of the curves reported in \figr{all_data}. Top panel: data of the
TN2mm prototype, represented with blue dots. Bottom panel: data of
the EM4mm, represented with red dots.   Error bars indicate only the
statistical uncertainty in the torque measurements.  The lines of
the radiation pressure contribution to the intercepts predicted by
different models are shown for comparison in both panels. The brown
continuous lines show the prediction of the infinite plates
approximation ($\frac{8\sigma}{3c}T_0^3$ as in \eqr{radiation:3}),
which is the same for both sensors. The other models are derived
from the numerical simulations discussed in \aref{rapress:app} and
using the temperature patterns estimated with the procedure
described in \aref{tempattern:app}. Different values have been taken
into account for the surface absorptivity $a$, and we indicate with
$d=100\%$ the case of diffuse light scattering and with $d=0\%$ the
specular light scattering case. Orange line: $a =100\%$, $d = 0\%$.
Black: $a = 10\%$, $d=100\%$. Magenta: $a=10\%$,$d=0\%$. Green: $a =
5\%$, $d=100\%$. Cian: $a = 10\%$, $d=0\%$.}
\end{figure}

\subsection{The zero pressure limit: thermal radiation pressure
and the effect of outgassing}

\begin{figure}[t]
\centering
\includegraphics[width=85mm]{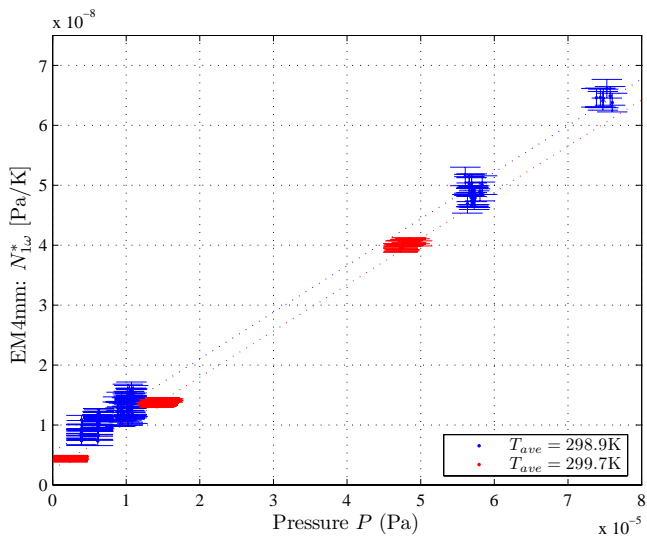}
\caption{\label{hysteresis} Two measurements of the thermal induced
normalized torque for the EM4mm sensor as a function of pressure, at
roughly the same average temperature (blue 298.9~K, red 299.7~K)
After the first measurement (blue), the sensor temperature was
progressively increased to 314~K (highest temperature in
\figr{all_data}) while measuring at the higher temperatures of the
study. The second (red) data set was measured subsequently after
cooling to 299.7~K. The slopes agree within the measurement
uncertainty after correction for the small temperature difference.
The measured intercept, on the other hand, is reduced by more than a
factor 2 in the second measurement, indicating a sharp reduction of
the outgassing effect during the ``bakeout'' at 314~K.}
\end{figure}

In \figr{interc} we plot the zero-pressure values of the normalized
torques, $N^*(P=0)$, extrapolated to $P=0$ from the data in
\figr{all_data}, as a function of the average temperature for the
two sensors. Shown for comparison are different estimates of the
radiation pressure contribution. These estimates, evaluated for the
experimental temperature distributions as estimated in Appendix C,
include the infinite plate model of \eqr{radiation:3} (brown), and
the results of the radiation pressure simulation presented in
\sref{rapress:app}) for 5\% absorption and both diffuse (green) and
specular (cian) reflection. As discussed in \sref{rapress:app}, for
the geometries of these GRS prototypes, the expected thermal
radiation effect is substantially suppressed, relative to the
infinite plate prediction, a suppression that is enhanced with
increasing reflectivity. Reflectivities of 95 \% ($a = 0.05$) or
higher are in fact expected for gold coating at reflectivity thermal
radiation wavelengths \cite{radiaprop}. While the data for the TN2mm
are consistent with the presence of the radiation pressure effect,
the data for the EM4mm prototype dip slightly below even the 95\%
reflection prediction for the radiation pressure effect at the
lowest temperatures. This discrepancy will be addressed shortly.

For both sensors, we observe an increase in the zero-pressure torque
intercepts with temperature, as expected both for the radiation
pressure and outgassing effects. However, for both sensors, the
observed increase with temperature is clearly faster than the $T^3$
dependence expected for radiation pressure, regardless of the
surface reflection properties or geometry, and suggests the
exponential-like dependence expected for outgassing. Further
evidence for the existence of outgassing-related phenomena is found
in the hysteretic behavior of the intercept following a thermal
cycle (see \figr{hysteresis}). This shows that even a mild
``bakeout'' at 314~K suppresses the intercept near 299~K by more
than a factor two, a hysteresis that must reasonably be associated
with the degassing of the surface at warmer temperatures.

While outgassing is thus the main effect contributing to the zero
pressure data, we note that the its magnitude near 293~K is, for the
TN2mm prototype, only roughly equal to that calculated for the
infinite plate model for the radiation pressure effect and, for the
EM4mm sensor designed for LISA Pathfinder, considerably smaller. As
such, the outgassing effect as measured here does not pose a
significant increase to the thermal gradient noise budget already
considered for LISA.

Even if radiation pressure plays a negligible role in the
zero-pressure torque, no simple model can explain the observation
that, for the EM4mm prototype, the extrapolated zero-pressure
intercept of the normalized torque becomes negative at the lowest
temperatures studied.  This small -- roughly 15\% of the magnitude
of the infinite plate prediction for the radiation pressure effect
-- but statistically significant violation of the expected sign for
a thermal gradient induced torque is visible in the 283~K data point
in \figr{interc}.  There are several possible explanations for this
observation:
\begin{itemize}
\item{ {\it Pressure offset}
If the ion gauge pressure measurements were higher, by a constant
offset $P_{OFF}$, than the actual pressure inside the sensor, then
the estimated zero-pressure torque intercepts would be artificially
lowered. This hypothesis is at least consistent with the fact that,
in spite of the fitted negative zero-pressure torque intercepts, we
do not observe negative torques even at the lowest measured pressure
values. $P_{OFF}$ would have to be roughly $2\times 10^{-6}$~Pa to
explain the observed negative torque offset. Given the quoted
accuracy of the Pfeiffer gauge, this offset should not exceed
$1.5\times 10^{-6}$~Pa (15\% of the $10^{-5}$ ~Pa full scale used
for the low pressure data). Such a pressure measurement offset could
thus explain most, but not all, of the observed negative torque
intercept. Another, slightly less stringent, upper limit on
$P_{OFF}$ comes from the lowest pressure measured in the study, $2.3
\times 10^{-6}$~Pa.

We also note that an offset arises in the outgassing from inside the
sensor, which causes the true pressure inside the sensor to exceed
that measured in the vacuum chamber. This thus has the wrong sign to
explain the measured negative zero-pressure torques. Additionally,
this effect is estimated to be quite small; given the effective
impedance of the holes connecting the sensor to the rest of the
vacuum chamber, the pressure inside the sensor would be only
10$^{-6}$~Pa higher than outside, even if as much as 10\% of the
total measured apparatus outgassing (roughly $3 \times 10^{-6}$~mbar
l / s) originates inside the GRS. }

\item {{\it Inhomogeneous outgassing}
Outgassing, associated with contamination and ``virtual leaks,''
such as the joints between different pieces, is likely to be highly
dependent on position inside the GRS. It is possible that, at least
for certain temperatures, the GRS outgassing could be dominated by
one or several localized points that are in positions, likely near
the corners of the sensor, that are warmed in phase with the driven
heaters during the experiment, but nonetheless contribute a torque
of the opposing sign. }
\end{itemize}
At present we can not discriminate between these or other
explanations for the negative zero-pressure torque observed at 283~K
for the EM4mm prototype.

\section{Conclusions}
\label{conclusions} This paper represents the first extensive study
aimed at characterizing thermal-gradient induced forces in the
specific environment relevant to LISA free-fall, realistically
capturing the key environmental parameters that the LISA test masses
will experience inside of the gravitational reference sensors in
orbit. This includes not only temperature and residual gas pressure,
but also geometry, materials, construction technique, and handling
history. The conclusions of the study, based on torsion pendulum
measurements made with two representative GRS prototypes and
numerical simulations of the radiometric and radiation pressure
effects, is that thermal gradient forces will likely not threaten
the LISA free-fall goals.

For the radiometric effect, the calculated thermal gradient transfer
function is roughly 25\% larger than that predicted by the simple
infinite plate formula. This increase -- yielding $\left
.\frac{dF_x}{d\Delta T} \right| _R \approx 23 \mathrm{pN /K}$ or
$\kappa_{R} \approx 1.25$ -- is due largely to the shear forces
exerted by molecules striking the $Y$ and $Z$ faces of the TM.  In
the experiments performed, the radiometric effect can be isolated
from the pressure dependence of the measured torques in the
measurements present here.  The expected $\frac{P}{T_0}$ functional
dependence, as well as a frequency-independent linear response, is
observed. The amplitude is quantitatively consistent with that
predicted by the simulations, for both sensor prototypes, but the
systematic uncertainty in the estimation of the relevant temperature
profile -- estimated to be roughly 20\% for the prototype more
closely representing the current LISA design (EM4mm) -- is not
sufficient to distinguish the corrections to the infinite plate
model foreseen in the simulations.

>From the radiation pressure simulations, we observe how larger gaps
allow a dilution of the force from bouncing thermal photons, given
the highly reflective GRS envisioned for LISA. Assuming 95 \%
reflectivity and specular reflection, the resulting radiation
pressure transfer function is reduced to roughly a third of the
infinite plate model prediction ($\kappa_{RP} \approx 0.32$),
yielding $\left . \frac{dF_x}{d\Delta T} \right|_{RP} \approx 9
\mathrm{pN /K}$. Radiation pressure, along with outgassing and any
other pressure-independent effect, are studied with the torque data
in the limit of zero pressure. These data here are dominated by the
outgassing effect, clearly distinguished from the radiation effect
by the sharp temperature dependence and the observed reduction in
the effect following a mild ``bakeout.'' However, the effect is not
large enough to threaten the LISA noise budgets; for both
prototypes, the torque at room temperature is comparable to (the
older 2 mm-gap prototype sensor) or smaller than (LISA 4 mm design)
that foreseen in the infinite parallel plate model for radiation
pressure. Additionally, it is likely that the outgassing effect,
already tolerable for LISA in these sensors which have never had a
serious bake-out, would very likely be further reduced with the
180~K bakeout foreseen before flight.

The measured torques are thus consistent with the predicted
radiometric effect and reveal no large zero-pressure effect, from
outgassing or anything else, that exceeds the levels already
included in the previously-assumed radiation pressure estimates. As
such, these experiments increase confidence in the LISA noise model
and suggest that the overall thermal gradient transfer function is
no larger than that indicated in the introduction,
$\frac{dF_x}{d\Delta T} \approx 100 \mathrm{pN / K}$. One caveat,
however, is that these experiments are based on torque measurements,
rather than the $x$ force relevant to LISA. Outgassing is not likely
to be homogeneously distributed within the sensor, and the torque
measurements are sensitive to most, but not all outgassing
locations, particularly those acting centrally on the TM face. To
address this concern, we are currently developing a new torsion
pendulum, with the TM displaced from the torsion axis, that will be
directly sensitive to the thermal gradient-induced $x$ forces
\cite{fourmasses}. In addition to a direct measurement of the
quantity of interest for LISA free-fall, the pendulum geometry,
sensitive to the translational temperature difference $\Delta T_x$
will allow a much simpler analysis of the relevant temperature
profile.

\appendix

\section{ Radiometer effect: numerical simulations details}\label{radiom:app}

The basic idea is to simulate a gas of non-interacting particles
bouncing in the gaps between the TM and the EH, in the presence of a
thermal gradient along the EH. The ballistic motion of a single
particle was calculated, with the impulse and moment of the impulse
recorded at every collision of the molecule with the TM. The total
TM force and torque exerted by the particle are calculated by
multiplying by the number of particles that would fill the gaps at a
given mean temperature and pressure.

\begin{figure}[t]
\includegraphics[width=70mm]{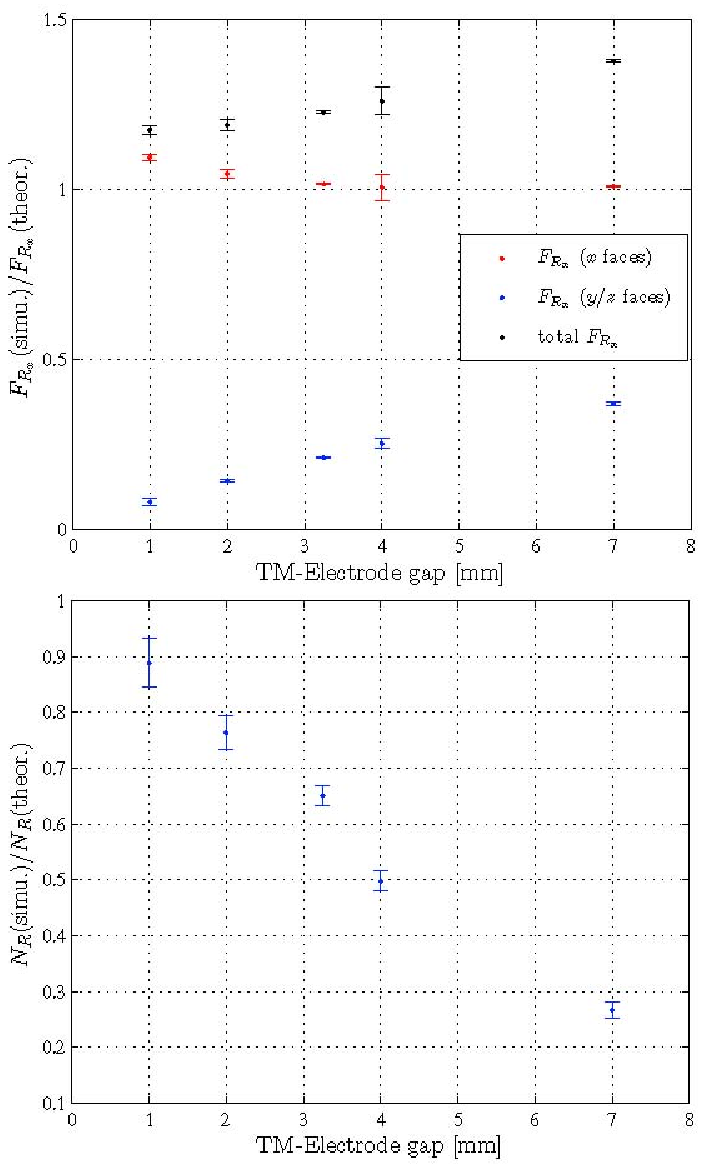}
\caption{\label{simula:radiom}Results of numerical simulations for
the radiometer effect, as a function of the gap (in mm) between test
mass and electrodes of the position sensor. The top panel shows the
ratio of the simulation torque to that calculated from
\eqr{Force:radiom}, for the $x$ component of the force acting on the
TM $X$ faces (red dots), for the $x$ force acting on the $Y$ and $Z$
TM faces (blue dots) and for the total $x$ force acting on the TM
(sum of previous values, shown as green dots). The bottom panel
shows the ratio of the simulated torque, around the $z$ (or $y$
axis, to the infinite plate result (\eqr{radiom:torqtheo}), for the
$Y$ (or $Z$) face. The average of the magnitudes for the 4 $X-Y$
faces has been reported. While all other calculations are made for
equal gaps in all directions, the point at 3.2 mm is calculated with
the real gaps of the configuration baselined both for LISA
Pathfinder and LISA and shown in \figr{sens}.}
\end{figure}

The sensor geometry considered is that of a cubic 46 mm TM placed in
the center of a rectangular box, both of which are fixed during the
simulation. There is no hole in the sensor that would allow exchange
of particles with the surroundings. As edge effects are expected to
be more important when the gap between the TM and the EH increases,
several configurations with different gaps has been analyzed,
including the design under construction for LISA Pathfinder and
currently proposed for LISA (this geometry is represented in
\figr{simula:radiom} by the point at 3.2 mm, which is the average of
the gaps along the $Y$ and $Z$ axes, 2.9 mm and 3.5 mm,
respectively). The gap sizes considered range from 1 mm to 8 mm.

The calculations for the translational force for LISA are performed
with a linear gradient across the position sensor along the $x$
axis, with the inner $X$ faces $X_{+}$, $X_{-}$ each at uniform
temperature and the inner surfaces of the $Y$ and $Z$ faces having a
linear temperature profile.  The TM is assumed to be isothermal at
the average temperature of the position sensor. We note that the
calculation, as it has been set up here, must be performed with a
temperature distribution defined in every point in the sensor, and
any modification of the temperature distribution, even locally, must
be studied as an entirely new simulation.

At every bounce onto a surface, the particle is assumed to
thermalize with the surface, with reemission that has no memory of
what happened before the bounce. When the particle leaves a surface,
we extract the velocity direction from a Knudsen distribution
\cite{Knudsen}, with flat probability both in the azimuthal angle
$\phi$ and in $\cos^2(\theta)$, where $\theta$ is the angle with
respect to the surface normal vector. The kinetic energy is
extracted from a distribution of velocities that would produce
Maxwell-Boltzmann distribution in the space.

Each time a particle hits a TM face, it completely transfers its
momentum to the TM. Upon reemission, it transfers the recoil
momentum to the TM and then flies at constant velocity until it
reaches another surface. Momenta and angular momenta transferred to
the TM are summed bounce by bounce, as well as the flying time
between bounces. At the end, the total linear and angular momenta
transferred to the TM are divided by the total elapsed time, giving
the average force and torque acting on the TM. This calculation has
been performed for each momenta component transferred to each TM
face. For each geometrical configuration, roughly 10 runs each
consisting of roughly $ 5\times 10^7$ bounces were performed.
 The resulting
force and torque values were averaged and the standard deviation
calculated to determine the calculation uncertainties.\\ The results
of the calculation are summarized in \figr{simula:radiom}.The GRS
design under construction for LISA Pathfinder and currently proposed
for LISA, is represented in \figr{simula:radiom} by the point at 3.2
mm. In \sref{radiom:theor}these results are illustrated and
discussed.

\section{Thermal radiation pressure: numerical
simulations}\label{rapress:app} The effective role of thermal
radiation pressure has been evaluated in the simplified GRS
geometry, with a cubic TM contained and centered inside a
rectangular box, by numerical simulations. The eventual
modifications to the radiation pressure ``transfer function'' can be
calculated and expressed as the correction factors $\kappa_{RP}$ and
$\gamma_{RP}$ introduced in \sref{rapress:theor}, for  comparison to
the simple model discussed there.

For the simulations, the total forces and torques on the TM
associated with thermal radiation have been represented as an
integral of $T^4$ multiplied by an independent vectorial force or
torque function, $\overrightarrow{f}(\mathbf{r})$ or
$\overrightarrow{n}(\mathbf{r})$, representing the contribution (per
K$^4$) of thermal photons originating from the sensor surface
element $d\mathbf{s}$ at position $\mathbf{r}$ of the EH surfaces
$\mathbf{S}$:
\begin{eqnarray}
\overrightarrow{F}_{RP}&=&\sum_{(\text{EH,TM faces})}\int
\overrightarrow{f}(\mathbf{r})T^4(\mathbf{r})d\mathbf{s}\nonumber\\
\overrightarrow{N}_{RP}&=&\sum_{(\text{EH, TM faces})}\int
\overrightarrow{n}(\mathbf{r})T^4(\mathbf{r})d\mathbf{s}\nonumber
\end{eqnarray}
In the infinite plate limit (see \eqr{radiation:2}), $f_x=\mp
\frac{2 \sigma}{3 c}$ for sensor surface elements directly opposite
the TM faces on the $X_+$ and $X_-$ faces (the yellow and orange
zones at left in \figr{radiom:torqtheo}), respectively, and zero
elsewhere in the sensor. For the torque, $n_z = \frac{2 \sigma}{3 c}
\times b(\mathbf{s})$, where the effective armlength $b(\mathbf{s})
= \pm y$ for the orange and yellow zones at right in
\figr{radiom:torqtheo} (and, analogously, $\pm x$ for the
corresponding zones on the $Y$ faces). The functions
$\overrightarrow{f}(\mathbf{r})$ and
$\overrightarrow{n}(\mathbf{r})$ depend only on the sensor geometry
and reflection properties, and so can be evaluated point-by-point
and independently of the sensor temperature distribution.

In the simulations, a chosen number of photons is emitted from a
given position of the TM or of the EH surfaces. Photon angle is
determined statistically, again assuming a uniform probability
density in $\phi$ and $\cos^2 \left( \theta \right)$. The surface
absorption properties were assumed to be independent of wavelength,
which allowed for a simple normalization for the number of photons
(or effective time of the simulation) and the resulting forces,
using the Stefan-Boltzmann law for the radiated power from a surface
element of given absorption coefficient $a$, $\left(a \sigma T^4
\right)$, and the energy - momentum relation for photons, $E = p c$.
Each emitted photon is propagated ballistically from the emitting
surface, and then either absorbed or reflected statistically,
according to the surface absorption properties imposed. In the case
of reflection, specular or diffuse (statistical angular reemission)
was decided statistically with the imposed percentage of diffuse
scattering, $d$. At each collision with the TM, the momentum and
moment of the momentum transfer are recorded and then summed at the
end of the simulation to determine the total force and torque on the
TM.

For the cases studied here, emission from the TM was not calculated,
as, in the isothermal TM-approximation reasonable for LISA and our
experiments, the TM radiation produces no net recoil. Additionally,
for simplicity, the absorption properties were assumed to be uniform
and equal for the TM and sensor surfaces. The points of emission for
which the simulation was performed were uniformly distributed on the
faces of the sensor, using a rectangular grid with $0.5$ mm spacing.
Each simulation, typically involving $200$ photons, was repeated,
typically $5$ times, for each point inside the sensor, in order to
obtain the uncertainties for $\overrightarrow{f}(\mathbf{r})$ and
$\overrightarrow{n}(\mathbf{r})$.

Simulations have been performed for several different sensor
geometries, including the two studied experimentally in this paper.
Additionally, simulations have been performed for several sets of
reflection properties: ($i$) absorptivity $a= 100\%$ ($ii$)
absorptivity = $a=10\%$ and pure diffusive reflection $(d=100\%)$
($iii$) absorptivity $a=10\%$ and pure specular reflection $(d=0\%)$
($iv$) absorptivity = $a=5\%$ and pure diffusive reflection
$(d=100\%)$ ($v$) absorptivity $a=5\%$ and pure specular reflection
$(d=0\%)$.

Finally, in order to obtain the relevant radiometric effect
correction factors for the $x$-force and the $z$-torque,
$\kappa_{RP}$ and $\gamma_{RP}$, the values of
$\overrightarrow{f}(\mathbf{r})$ and
$\overrightarrow{n}(\mathbf{r})$ obtained from the simulation must
be multiplied by $\left( T(\mathbf{r}) \right)^4$, with the sensor
temperature distribution $T(\mathbf{r})$ chosen to represent a
specific experimental thermal perturbation.  Results for cases
relevant to thermal gradient forces and torques are discussed below.

\subsection{Radiation pressure simulation results: induced force}
\label{sim_radiat_force} 
\begin{figure}
\centering
\includegraphics[width=80mm]{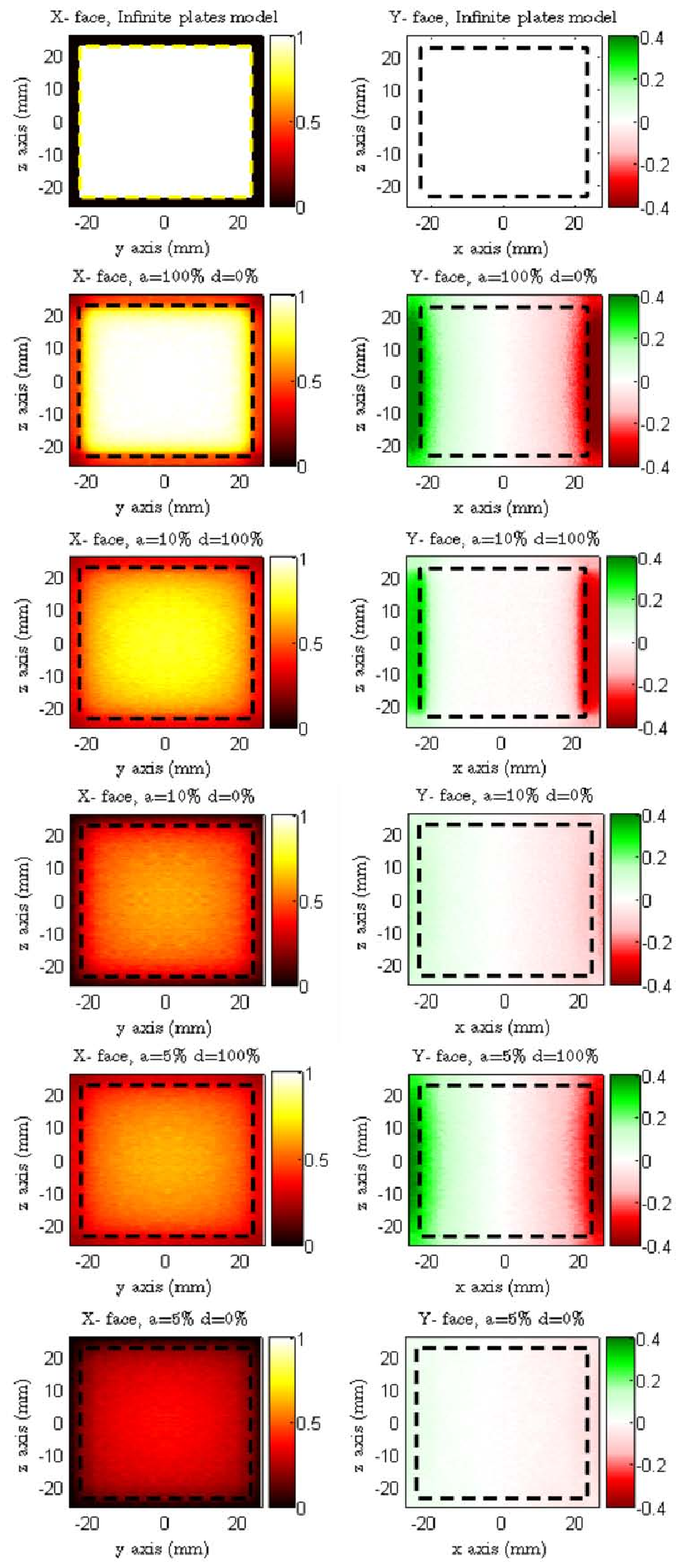}
\caption{Intensity plots of the $x$ component of the force function
$f_x(\mathbf{r})$, the force per K$^4$ per unit area of emitting
sensor surface $K^4$, normalized by $\left(2\sigma/3c\right)$, the
characteristic value of the infinite model. Results are shown for
emitting points on the $X$ (left column) and $Y$ surfaces (right
column), with the outline of the TM section a dashed-line 46 mm
square in each case. The $1^{st}$ row shows the data for the
infinite plates model. Rows 2 to 6 show the different cases of
reflection properties: full absorption, $a=100\%$, absorption
coefficient $a=10\%$ or $a=5\%$, pure diffusive reflection
($d=100\%$) or purely specular reflection ($d=0\%$). Legends for the
intensity values are shown in the colors bars on the right of each
plot.} \label{radpres:EM:force}
\end{figure}

For the study of the radiation pressure
force along $x$, we plot first the relevant force function
$f_x(\mathbf{r})$, normalized by the infinite plate value
$\left(2\sigma /3c\right)$, as a function of the emission position
$\mathbf{r}$ on the sensor $X$ and $Y$ faces (the $Z$ faces play a
role very analogous to that of the $Y$ faces for the $x$ force).
These are shown for the LISA Pathfinder sensor design in
\figr{radpres:EM:force} ($X$ faces - left column, $Y$ faces - right
column), together with infinite plates model case. Then, to
calculate the relevant factor $\kappa_{RD}$, integration of $f_x$ is
performed for a temperature distribution with a gradient in the $x$
direction, with each $X$ face at a uniform temperature and a linear
temperature gradient across the $Y$ and $Z$ faces.  The resulting
values for the correction factor $\kappa_{RP}$ are summarized for
different sensor geometries and absorptivity characteristics in
\tabr{radpress:tab:alpha}. The table also gives a breakdown of the
contributions to $\kappa_{RP}$, including not only the contribution
expected from the $X$ faces directly opposite the TM
($\kappa_{X(IN)}$), but also those not foreseen by the infinite
plate model, from the borders of the $X$ sensor faces
($\kappa_{X(OUT)}$) and from the other sensor faces ($\kappa_{YZ}$).

As visible in \figr{radpres:EM:force}($2^{nd}$ row, left), where we
consider full absorption of emitted photons ($a= 100\%$), for
emission from near the center of the sensor $X$ faces,
$f_x/\left(2\sigma/3c\right)\approx 1$, as expected for the infinite
plate model ($1^{st}$ row left). Approaching and crossing the edge
of the TM, $f_x/\left(2\sigma/3c\right)$ decreases smoothly from one
towards zero, as a continuously smaller fraction of the emitted
photons strike the TM surface. For runs with the smallest gaps (not
plotted here), the transition near the edge of the TM becomes
sharper, and the infinite plate model becomes more accurate.
Additionally, there is a significant contribution of emission from
the $Y$ faces to $f_x$ \figr{radpres:EM:force}(right column). The
zones near the edges adjacent to the $X_+$ and $X_-$ faces
contribute photons that are predominantly absorbed on the TM near
the center of the $Y$ face, and as such exert a shear force along
$x$. There is an analogous contribution from the $Z$ face. When
integrated with the temperature distribution, this shear force
contribution is largely responsible for the increase in the
radiation pressure effect ($\kappa_{RP} = 1.17$,
\tabr{radpress:tab:alpha}). We note that this increase in the case
of full absorption is similar to that seen for the radiometric
effect, with the shear forces along the $Y$ and $Z$ faces a leading
factor in both cases. Also in \tabr{radpress:tab:alpha}, we see that
with 200 micron gaps, the simulation confirms the infinite plate
model, $\kappa_{RP}=1$ to better than 1\%, nearly entirely from the
contribution of the $X$ faces opposite the sensor $\kappa_{x(IN)}$.

With higher reflectivity, i.e. low $a$ ($3^{rd}$ to $6^{th}$ row,
left), the effect of smoothing the $X$ face contribution is more
pronounced, as even photons emitted near the center of the $X$ face
can migrate off the TM $X$ face and deposit their momentum on other
faces, even on the opposite TM $X$ face. The smoothing is more
pronounced for specular scattering than diffuse, because in diffuse
scattering the effective distance traveled by the photons increases
only as the square root of the number of bounces, in a random walk,
rather than linearly as for specular reflection. The extreme case of
high reflectivity and specular reflection ($a=5\%$ and $d=0\%$,
bottom left) represents a progressive homogenization of the
radiation pressure, as a photon generated from any point in the
sensor winds up imparting momentum on different parts of the TM,
effectively washing out also the shear effect (see
\figr{radpres:EM:force}, bottom right). As a value for adsorption
around $5\div10\%$ and specular reflection is the likely case for
the gold coated surfaces envisioned for the LISA GRS, a reasonable
estimate of the radiation pressure correction for LISA ranges
between $\kappa_{RP} = \left(0.32 \pm 0.01\right)$ and $\kappa_{RP}
= \left(0.63 \pm 0.02\right)$ (the errors given are simply the
statistical uncertainties of the simulation).

\begin{table}[t]

\begin{ruledtabular}
\begin{tabular}{c| c| c c c c}
GRS Geometry    &    (absortivity,
& $\kappa_{RP}$  & $\kappa_{X(IN)}$   &$\kappa_{X(OUT)}$ &    $\kappa_{YZ}$  \\
(TM , gap) & diffusion)       &        &          &
     &              \\
\hline \hline
  &  (100\%,0\%)        & 1.17        &  0.95         &  0.05        &
0.17        \\
                     & (10\%, 100\%)   & 1.01       &
0.69      &   0.05     &   0.28         \\
(46mm$^3$, 4mm)                    & (10\%, 0\%)         & 0.63   &
0.50       &  0.02
&   0.10        \\
                    & (5\%, 100\%)         & 0.75   &  0.49       &
0.03       &   0.23        \\
                    & (5\%, 0\%)         & 0.32   &  0.26       &  0.01
&   0.05        \\
\hline
   &  (100\%,0\%)        &   1.12  &  0.96        &   0.04      &
0.12      \\
                     & (10\%, 100\%)   &  1.01  &
0.78      &   0.03       &   0.21           \\
  (40mm$^3$, 2mm)       & (10\%, 0\%)         & 0.75     &  0.64       &
0.02     &       0.10    \\
                 & (5\%, 100\%)         & 0.85     &  0.62       &  0.03
&       0.20    \\
                 & (5\%, 0\%)         & 0.45     &  0.39       &  0.01
&       0.05    \\
\hline \hline
&  (100\%,0\%)      &   0.999     &  0.986        &   0.002  &  0.01       \\
(46mm$^3$, 0.2mm)                 & (10\%, 100\%)   &  0.966     &
0.937        &   0.002 &   0.027        \\
                 & (10\%, 0\%)         & 0.946     &  0.906         &
0.002 &   0.038       \\
\hline
                   &  (100\%,0\%)      &   100  &  0.985        &   0.002
&  0.013       \\
 (40mm$^3$, 0.2mm)                & (10\%, 100\%)   &  0.963     &
0.929        &   0.002 &   0.031        \\
                    & (10\%, 0\%)         & 0.942     &  0.896         &
0.002 &   0.043       \\
\end{tabular}
\end{ruledtabular}
\caption{ \label{radpress:tab:alpha}The overall correction factor
$\kappa_{RP}$ estimated in the case of a linear temperature profile
across the sensor along x axis, and the X+ and X- faces at uniform
temperature;$\kappa =
\left(\kappa_{x_{in}}+\kappa_{x_{out}}+\kappa_{(y+z)}\right)$ and
$\kappa_{x_{in}}$, $\kappa_{x_{out}}$ and $\kappa_{(y+z)}$
representing respectively the contributions from the $X$ faces
facing the TM, the $X$ faces not facing the TM and the lateral $Y$
and $Z$ faces. The simulations were performed for different GRS
geometries (i.e. different gap); the one indicated with  (46mm$^3$,
4mm) is the GRS geometry under construction for  LISA Pathfinder and
currently baselined for LISA. Statistical errors in each estimate of
$\kappa_{RP}$ are in the order of $2\div4\%$.}
\end{table}

\subsection{Radiation pressure simulation results: induced torques}
\label{sim_radiat_torque}

\begin{figure}[t]
\centering
\includegraphics[width=85mm]{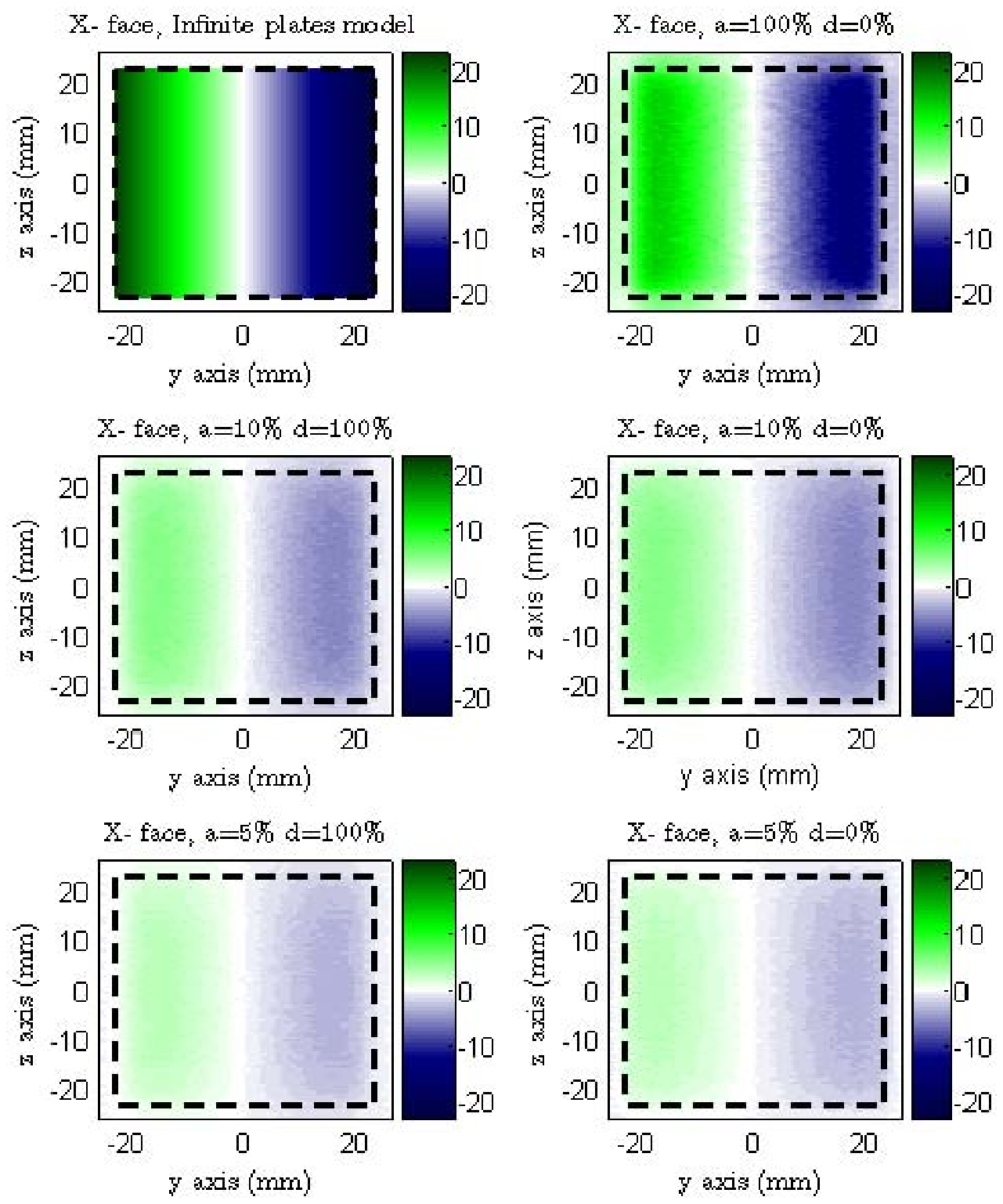}
\caption{\label{radpres:EM:torq} Intensity plot of the torque per
unit area $n_z(\mathbf{r})$ about the $z$ axis, normalized here by
the factor $\left(2\sigma /3c\right)$, for emitting points on the
$X$ surface (the $y$ face of the EH shows the exact same profile).
the $1^{st}$ plot (top left) shows the infinite plates model case.
The various cases of full absorption, $a=100\%$, absorption
coefficient $a=10\%$ or $a=5\%$, pure diffusive reflection
($d=100\%$) or purely specular reflection ($d=0\%$) are shown.}
\end{figure}
In \figr{radpres:EM:torq} we show the $z$ components
$n_z(\mathbf{r})$ of the torques per unit area per K$^4$ as a
function of the position inside the EH and as a function of the GRS
surface properties. We still use the normalization factor
$\left(2\sigma /3c\right)$:  we thus remark that
$n_z(\mathbf{r})/\left(2\sigma /3c\right)=b(\mathbf{r})$ for the
infinite plate approximation (shown in \figr{radpres:EM:torq} top
left), which coincides in the particular case of the $X$ face of the
EH under consideration with the $y$ coordinate, and with the $x$
coordinate on the $Y$ face. The different cases of surface
reflectivity are here shown (left to right, top to bottom).

As is visible,  a reduction similar to that observed for the force
is seen for the thermal radiation torque. However, in this case, the
effect is much more relevant: the overall attenuation of $N_z$ is
much more effective, because the attenuation of $f_z(\mathbf{r})$ is
progressively more effective approaching the edges of the TM. It
thus gets reduced exactly where the arm-length $b(\mathbf{r})$ is
bigger and the torque per unit area should contribute most to the
overall effect. This is clearly visible in the data in
\tabr{radpress:tab:beta}, where we report the estimates of the
radiation pressure induce torque correction factors $\gamma_{RP}$,
for an ideal linear temperature profile on one of the faces of the
IS (with zero temperature modulation on the other faces). We
estimate the total torque to be reduced down to $\approx 73\%$
compared to the theoretical estimate in the case of $100\%$
absortivity. In the case of absortivity of the IS surfaces of order
$10\%$ or lower, the reduction is then much more relevant, and we
estimate the total torque to be suppressed down to a value between
$15 \%$ for EM4mm and $24\%$ for TN2mm of the theoretical value, in
the realistic case of reflectivity $95\%$ ($a=5\%$) and pure
specular reflection ($d=0\%$)(see \tabr{radpress:tab:beta}).

\begin{table}[th]
\begin{tabular}{c| c| c }
\hline \hline
GRS Geometry    &    (absorptivity,           & $\gamma_{tot}$    \\
(TM , gap) & diffusion)       &         \\
\hline \hline
  &  (100\%,0\%)        & 0.73               \\
(46mm$^3$, 4mm)                     & (10\%, 100\%)   & 0.30              \\
                    & (10\%, 0\%)         & 0.27        \\
                    & (5\%, 100\%)         & 0.17          \\
                    & (5\%, 0\%)         & 0.15          \\
\hline
   &  (100\%,0\%)        &   0.80        \\
(40mm$^3$, 2mm)                     & (10\%, 100\%)   & 0.42          \\
                 & (10\%, 0\%)         & 0.39        \\
                 & (5\%, 100\%)         & 0.27       \\
                 & (5\%, 0\%)         & 0.24       \\
\hline \hline
\end{tabular}
\caption{ \label{radpress:tab:beta} Correction factors $\gamma_{RP}$
for the thermal radiation pressure induced torques. The GRS
geometries here reported (46mm$^3$, 4mm and 40mm$^3$, 2mm)
correspond to the two GRS prototypes studied with the torsion
pendulum (EM4mm and TN2mm respectively). Statistical errors are in
the order of $2\div4\%$.}
\end{table}

\section{Estimate of temperature patterns and thermal integrals}
\label{tempattern:app}

As mentioned in \sref{meas_tech}, for each measurements cycle we
evaluate the component at frequency $\omega$ of the ``thermal
torque'' $\left( \int_S T(s) b(s) ds \right)_{(\omega)}$ with the
equation
\begin{eqnarray}
\left(\int_S T(s) b(s) ds\right)_{\omega} & = &
   \sum_{X_{+},X_{-}}
\int_A (\pm)  T(y,z,\omega) y dy dz + \nonumber\\
& + & \sum_{Y_{+},Y_{-}}  \int_A  \left( \mp \right) T(x,z,\omega) x
dy dz.
\end{eqnarray}
The components $T(s_{1},s_{2},\omega)$ of the temperature at
position$(s_{1},s_{2})$, where $s_1$ and $s_2$ are two coordinates
mapping the 4 $X$-$Y$ sensor surfaces, have been estimated by a
cubic spatial interpolation of the components
$T(s_{1},s_{2},\omega)_i$ of the readings $T_i(t)$ of the
thermometers on the external electrode housing surface at the
positions $(s_1,s_2)_i$, which were measured during assembly.  In
performing the interpolation, several important approximations were
made and are addressed in the next two paragraphs.

Figure \ref{therm:loc} shows a cartoon of the two GRS prototypes
with the locations of the thermometers (small blue squares). For the
EM4mm prototype, six of the eight installed thermometers were used
for the interpolation, namely $T1$, $T2$, $T3$, $T4$, $T7$, and
$T8$. For TN2mm, four of the five thermometers were used for the
interpolation ($T2$, $T3$, $T5$ and $T6$). In this case, the reading
of $T5$ was assigned to the temperature at the center of the heater
H1E, in order to constrain the local temperature maximum to coincide
with the heater location. Additionally, in order to construct the
temperature distribution on the TN2mm $X+$ face, which has only a
single thermometer, in contrast with the three $X-$ face
thermometers, we introduced an artificial thermometer reading,
$T_{fake}$ near the heater H2W. The modulation of $T_{fake}$ was
assigned to be the value of the interpolated component at the center
of the electrode below H2E on the opposing $X-$ face, scaled by the
ratio $\frac{T_2}{T_4}$, in order to account for observed asymmetric
heating for the two $X$ faces.

\begin{figure}[t]
\includegraphics[width=80mm]{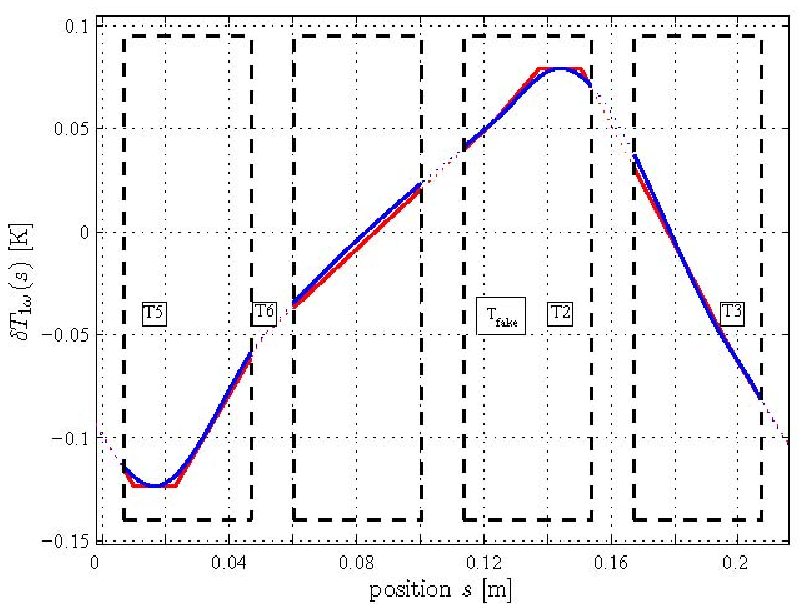}
\caption{\label{TN_temp_profiles} An example of the temperature
profile across the entire sensor TN2mm. Blue line: derived from
cubic interpolation of the thermometer readings. Red line: based on
a constant temperature across the electrodes with operating heaters
and a linear interpolation of the thermometer readings across the
rest of the sensor. A single temperature is assigned for all values
of $z$ at each given value the horizontal coordinate $s$ ($x$ or
$y$, depending on the face).}
\end{figure}

Figure \ref{TN_temp_profiles} shows an example of a reconstructed
temperature profile $T(s_{1},s_{2},\omega)$ across the entire sensor
TN2mm. Several assumptions were made in making this estimation, for
both sensors. First, the temperature on the inner surface of the
sensor is assumed to be equal to that on the corresponding outer
surface position, as calculated by the interpolated the outer
surface thermometer readings. Additionally, the temperature profile
along the $z$ axis (that parallel to the torsion fiber in our
experiments) was collapsed to a single value, that of the
temperature readings interpolated in the horizontal dimension. The
measured temperatures are thus assumed to represent the average
temperatures along the sensor $z$ axis at a given value of the
horizontal coordinate $x$ (or $y$, depending on the face). These
assumptions clearly result in some systematic error in the
estimation of the thermal integral, which is discussed next.

The temperature read by a thermometer on the GRS external surface is
a good approximation of that measured on corresponding the inner GRS
surface position, as long as the thermometer is not too close to an
operating heater or to relevant localized thermal interfaces, such
as the supporting legs at the four lower corners of the sensor. All
the thermometers used in the interpolation are indeed far from these
corner supporting legs, and, additionally, they are all placed at
least 9~mm from the border of the heaters, a distance equal to the
sensor wall thickness, including the electrode. A simplified FEM
model has confirmed that the inside-outside temperature difference
can indeed be considered negligible in these conditions.

As stated above, the thermometer readings used for the interpolation
are assumed to be the average temperature value along $z$. For the
$Y$ faces, where there are no heaters, the main effect of the
temperature variation along $z$ is expected to be an
altitude-dependent offset of the whole profile, with a negligible
impact on the thermal integral $\int_S T(\mathbf{s})
b(\mathbf{s})ds$, where a constant temperature offset integrates to
zero. For the $X$ faces, however, the operating heaters produce a
more complicated and more rapidly varying temperature pattern.
Moreover, the inner surface temperature modulation in correspondence
with the operating heater is expected to be relatively uniform
across the heater, but higher than that measured by the closest
thermometer. A rough estimation of the consequent correction to the
temperature pattern on the $X$ faces has been obtained by assuming
that the heat flux in the proximity of a cylindrical heater is
basically radial and uniform, an approximation supported by a
simplified FEM calculation. In this approximation, we estimate that
our simple interpolation technique underestimates the thermal
integral by roughly $30\%$ for the UTN2mm and $20\%$ for the EM4mm.
However, given the approximate nature of the analysis and the
experimental difficulty of verifying this model with a tight array
of thermometers on the inner sensor surfaces, we choose here not to
correct the data for this likely systematic error. Instead, we
present the data obtained from the relatively straightforward
interpolation and discuss the role of this systematic error in the
conclusion of our data analysis.

In order to address the dependence of the thermal integral
calculation on the method of spatial interpolation, we compare the
results of two different methods, both of which reasonably reproduce
the expected relatively uniform temperature profile of the ``hot''
electrodes coinciding with the locations of active heaters (see
\figr{TN_temp_profiles}). The first calculation assumes a perfectly
constant temperature across the ``hot'' electrodes and a linear
interpolation for the rest of the temperature profile. These results
are compared with that obtained with the cubic polynomial
interpolation. Thanks to the higher number of thermometers, for the
EM4mm prototype the difference was negligible, while for the TN2mm
model the first method gave a value $20\%$ larger than the second.

Considering the approximate analysis of the different sources of
uncertainty in the thermal integral, we conservatively assume the
overall uncertainty to be the linear sum of the above estimated
errors. This yields an uncertainty of roughly $50\%$ for the TN2mm
prototype and roughly $20\%$ for EM4mm. Given that the same heating
pattern is repeated in all measurements presented for each of the
two sensors, this uncertainty in the thermal integral estimation can
be thought of as a single scale factor for the each of the two
sensor data sets, rather than an uncertainty acting incoherently on
different data points in the same sensor. Unless otherwise
specified, the error bars of the torque data normalized by means of
the "thermal integral", namely $N^*$ represents just the statistical
error of the torque measurement $N$. The role of this scale factor
will be discussed with the conclusion of our data analysis in
\sref{exp_results_disc}.

\begin{acknowledgements}
The authors would like to thank Albrecht R\"udiger and Peter Bender
for bringing thermal gradient noisy forces to our attention and for
many useful discussion and Paolo Bosetti for help with the FEM
modeling. This work was supported by ASI.

\end{acknowledgements}


\begin{thebibliography}{20}
\expandafter\ifx\csname
natexlab\endcsname\relax\def\natexlab#1{#1}\fi
\expandafter\ifx\csname bibnamefont\endcsname\relax
  \def\bibnamefont#1{#1}\fi
\expandafter\ifx\csname bibfnamefont\endcsname\relax
  \def\bibfnamefont#1{#1}\fi
\expandafter\ifx\csname citenamefont\endcsname\relax
  \def\citenamefont#1{#1}\fi
\expandafter\ifx\csname url\endcsname\relax
  \def\url#1{\texttt{#1}}\fi
\expandafter\ifx\csname urlprefix\endcsname\relax\def\urlprefix{URL
}\fi \providecommand{\bibinfo}[2]{#2}
\providecommand{\eprint}[2][]{\url{#2}}

\bibitem[{big()}]{bender:LISA}
\bibinfo{note}{P. Bender et al, LISA ESA-SCI(2000)11, 2000.}

\bibitem{GPB} {http://einstein.stanford.edu/, http://www.gravityprobeb.com}     

\bibitem{STEP} {STEP Satellite Test of the Equivalence Principle, Report on     
the Phase A Study, ESA/NASA-SCI(93)4, 1993.}

\bibitem{Microscope} {www.cnes.fr/activities/connaissance/ physique/microsatellite/}  

\bibitem[{\citenamefont{Dolesi et~al.}(2003)}]{dolesi:sensor}
\bibinfo{author}{\bibfnamefont{R.}~\bibnamefont{Dolesi}} \bibnamefont{et~al.},
  \bibinfo{journal}{Class. Quant. Grav.} \textbf{\bibinfo{volume}{20}},
  \bibinfo{pages}{S99} (\bibinfo{year}{2003}).

\bibitem[{\citenamefont{Stebbins et~al.}(2003)}]{stebbins:errors}
\bibinfo{author}{\bibfnamefont{R.T.}~\bibnamefont{Stebbins}} \bibnamefont{et~al.},
  \bibinfo{journal}{Classical and Quantunm Gravity} \textbf{\bibinfo{volume}{21}},
  \bibinfo{pages}{S653} (\bibinfo{year}{2004}).

\bibitem[{\citenamefont{Schumaker et~al.}(2003)}]{bonnie:noise}
\bibinfo{author}{\bibfnamefont{B.L.}~\bibnamefont{Schumaker}},
  \bibinfo{journal}{Classical and Quantum Gravity} \textbf{\bibinfo{volume}{20}},
  \bibinfo{pages}{S239} (\bibinfo{year}{2003}).



\bibitem[{\citenamefont{Weber et~al.}(2002{\natexlab{b}})}]{weber:sensor}

\bibinfo{author}{\bibfnamefont{W.~J.} \bibnamefont{Weber}}
  \bibnamefont{et~al.}, \bibinfo{journal}{SPIE Proc.}
  \textbf{\bibinfo{volume}{4856}}, \bibinfo{pages}{31}
  (\bibinfo{year}{2002}{\natexlab{b}}).



\bibitem[{\citenamefont{Hueller et~al.}(2005)}]{mauro:mag}
\bibinfo{author}{\bibfnamefont{M.}~\bibnamefont{Hueller}
},\bibnamefont{et~al.},
  \bibinfo{journal}{Classical and Quantum Gravity} \textbf{\bibinfo{volume}{22}},
  \bibinfo{pages}{S521} (\bibinfo{year}{2005}).

\bibitem{vitale:nuc:phys} S. Vitale et al., Nuclear Physics {\bf B}, v.110, p.209, 2002

\bibitem[{\citenamefont{Anza et~al.}(2005)}]{vitale:LTP}
\bibinfo{author}{\bibfnamefont{S.}~\bibnamefont{Anza}} \bibnamefont{et~al.},
  \bibinfo{journal}{Class. Quant. Grav.} \textbf{\bibinfo{volume}{22}},
  \bibinfo{pages}{S125} (\bibinfo{year}{2005}).

\bibitem[{\citenamefont{Hueller et~al.}(2002)}]{hueller:pend}
\bibinfo{author}{\bibfnamefont{M.}~\bibnamefont{Hueller}} \bibnamefont{et~al.},
  \bibinfo{journal}{Class. Quant. Grav.} \textbf{\bibinfo{volume}{19}},
  \bibinfo{pages}{1757} (\bibinfo{year}{2002}).

\bibitem[{\citenamefont{Carbone et~al.}(2007)}]{prd:force_noise}
\bibinfo{author}{\bibfnamefont{L.}~\bibnamefont{Carbone}} \bibnamefont{et~al.},
  \bibinfo{journal}{Phys. Rev. D} \textbf{\bibinfo{volume}{75}},
  \bibinfo{pages}{042001} (\bibinfo{year}{2007}).

\bibitem[{\citenamefont{Carbone et~al.}(2003)}]{carbone:prl}
\bibinfo{author}{\bibfnamefont{L.}~\bibnamefont{Carbone}} \bibnamefont{et~al.},
  \bibinfo{journal}{Phys. Rev. Lett.} \textbf{\bibinfo{volume}{91}},
  \bibinfo{pages}{151101} (\bibinfo{year}{2003}).


\bibitem[{\citenamefont{Carbone et~al.}(2003)}]{hueller:upperlimits}
\bibinfo{author}{\bibfnamefont{L.}~\bibnamefont{Carbone}} \bibnamefont{et~al.},
  \bibinfo{journal}{Class. Quant. Grav.} \textbf{\bibinfo{volume}{21}},
  \bibinfo{pages}{S611} (\bibinfo{year}{2004}).

\bibitem{carbone:thesis} {L.Carbone, {\it Ground based investigation of force noise sources for LISA}, PhD Thesis, University of
Trento, Feb. 2005}
\bibitem[{\citenamefont{Carbone et~al.}(2005)}]{carbone:char}
\bibinfo{author}{\bibfnamefont{L.}~\bibnamefont{Carbone}} \bibnamefont{et~al.},
  \bibinfo{journal}{Class. Quant. Grav.} \textbf{\bibinfo{volume}{22}},
  \bibinfo{pages}{S509} (\bibinfo{year}{2005}).



\bibitem{seattle} \bibinfo{author}{S. Schlamminger et~al}, \bibinfo{journal}{Proceedings of the 6th International
LISA Symposium, AIP Conf. Proc.},\textbf{\bibinfo{volume}{873}},
\bibinfo{pages}{151-157} (\bibinfo{year}{2006}).

   \bibitem[{\citenamefont{Nobili et~al.}(2001)}]{nobili}
\bibinfo{author}{\bibfnamefont{A.M.}~\bibnamefont{Nobili}} \bibnamefont{et~al.},
  \bibinfo{journal}{Phys. Rev. D} \textbf{\bibinfo{volume}{63}},
  \bibinfo{pages}{101101(R)} (\bibinfo{year}{2001}).

 \bibitem[{\citenamefont{Nobili et~al.}(2002)}]{nobili_astr}
\bibinfo{author}{\bibfnamefont{A.M.}~\bibnamefont{Nobili}} \bibnamefont{et~al.},
  \bibinfo{journal}{New Astronomy} \textbf{\bibinfo{volume}{7}},
  \bibinfo{pages}{521} (\bibinfo{year}{2002}).

\bibitem{pollack:thermal} \bibinfo{author}{S. E. Pollack et~al.}, \bibinfo{journal}{Proceedings of the 6th international
LISA Symposium, AIP Conf. Proc.},\textbf{\bibinfo{volume}{873}},
\bibinfo{pages}{158-164} (\bibinfo{year}{2006}).


\bibitem{Bender} \bibinfo{author}{P.L. Bender}, \bibinfo{journal}{Proceedings of the 6th international
LISA Symposium, AIP Conf. Proc.},\textbf{\bibinfo{volume}{873}},
\bibinfo{pages}{143-150} (\bibinfo{year}{2006}).

\bibitem{Merko} \bibinfo{author}{S.M. Merkowitz}, \bibinfo{journal}{Proceedings of the 6th international
LISA Symposium, AIP Conf. Proc.},\textbf{\bibinfo{volume}{873}},
\bibinfo{pages}{133-142} (\bibinfo{year}{2006}).

\bibitem{Roth:book}{A. Roth , Vacuum Technology, North Holland (1982)}

\bibitem[{\citenamefont{Wu Y. et~al.}(2002)}]{Wu}
\bibinfo{author}{\bibfnamefont{Y.}~\bibnamefont{Wu}},
  \bibinfo{journal}{J. Vac. Sci. Technol.} \textbf{\bibinfo{volume}{9}},
  \bibinfo{pages}{1248} (\bibinfo{year}{1972}).

\bibitem[{\citenamefont{Wu}(2002)}]{Wu2}
\bibinfo{author}{\bibfnamefont{Y.}~\bibnamefont{Wu}},
  \bibinfo{journal}{J. of Chem. Phys.} \textbf{\bibinfo{volume}{48}},
  \bibinfo{pages}{889} (\bibinfo{year}{1968}).

\bibitem {radiaprop}
  \bibinfo{journal}{Thermal Radiative Properties: metallic elements
  and alloys,Thermophysiscal prop. of Matter, the TPRC Data Series} \textbf{\bibinfo{volume}{7}},
  \bibinfo{pages}{244-245}.


\bibitem[{big()}]{vitale:therm:simu}
\bibinfo{note}{S. Vitale, Unitn-Int-1-2003, 2003.}


\bibitem{rudiger} A. R\"{u}diger, {\it Residual gas effects in space-borne position sensors},
 MPI f\"{u}r Gravitationsphysik, AEI Hannover, unpublished,
(2002)

\bibitem{hanlon:book}{ J.F.O'Hanlon, A User Guide to Vacuum Technology, Wiley-Interscience Publication(1989)}


\bibitem{Knudsen} \bibinfo{author}{W.Steckelmacher and M.W.Lucas},\bibinfo{journal}{ J.Phys.D: Appl.Phys.},\textbf{\bibinfo{volume}{16}},
\bibinfo{pages}{1453-1460} (\bibinfo{year}{1983}).


\bibitem{fourmasses}\bibinfo{author}{L. Carbone et al}, \bibinfo{journal}{Proceedings of the 6th international
LISA Symposium, AIP Conf. Proc.},\textbf{\bibinfo{volume}{873}},
\bibinfo{pages}{561-565} (\bibinfo{year}{2006}).

\end{thebibliography}
\end{document}